\documentclass[review]{elsarticle}

\usepackage{lineno,hyperref}
\modulolinenumbers[5]

\usepackage[nolist]{acronym}
\begin{acronym}
\acro{adam}[Adam]{Adaptive Moment Estimation}
\acro{ann}[ANN]{Artificial Neural Network}
\acro{cca}[CCA]{Canonical Correlation Analysis}
\acro{de}[DE]{Differential Evolution}
\acro{es}[ES]{Evolutionary Strategies}
\acro{fnn}[FNN]{Feedforward Neural Network}
\acro{fptz}[FPTZ]{Fastest Path to Zero Initiative}
\acro{ga}[GA]{Genetic Algorithms}
\acro{inl-nric}[INL-NRIC]{Idaho National Labs - Nuclear Reactor Innovation Center}
\acro{lr}[LR]{Linear Regression}
\acro{mds}[MDS]{Multi-Dimensional Scaling}
\acro{moo}[MOO]{Multi-objective Optimization}
\acro{nuram}[NURAM]{Nuclear Reactor Analysis and Methods}
\acro{neorl}[NEORL]{NeuroEvolution Optimisation with Reinforcement Learning}
\acro{neup}[NEUP]{Nuclear Energy University Program}
\acro{ns}[NS]{Non-Dominated Sorting}
\acro{nsgaii}[NSGA-II]{Non-dominated Sorting Genetic Algorithm II}
\acro{nsgaiii}[NSGA-III]{Non-dominated Sorting Genetic Algorithm III}
\acro{rfr}[RFR]{Random Forest Regression}
\acro{pca}[PCA]{Principal Component Analysis}
\acro{pso}[PSO]{Particle Swarm Optimization}
\acro{svi}[SVI]{Social Vulnerability Index}
\acro{relu}[ReLU]{rectified linear unit}
\acro{sa}[SA]{Simulated Annealing}
\acro{um}[UM]{University of Michigan}
\acro{c2n}[C2N]{Coal to Nuclear Transition}
\acro{stand}[STAND]{Siting Tool for Advanced Nuclear Development}
\acro{npp}[NPP]{nuclear power plant}
\acro{cpp}[CPP]{coal power plant}
\acro{epa}[EPA]{Environmental Protection Agency}
\acro{acres}[ACRES]{Assessment Cleanup and Redevelopment Exchange System}
\acro{smr}[SMR]{small modular reactor}
\acro{ga}[GA]{Genetic Algorithm}
\acro{es}[ES]{Evolutionary Strategies}
\acro{nsgaii}[NSGA-II]{Non-dominated Sorting Genetic Algorithm II}
\acro{nsgaiii}[NSGA-III]{Non-dominated Sorting Genetic Algorithm III}
\acro{ns}[NS]{Non-Dominated Sorting}
\acro{pca}[PCA]{Principal Component Analysis}
\acro{cca}[CCA]{Canonical Correlation Analysis}
\acro{moo}[MOO]{multi-objective optimization}
\acro{frs}[FRS]{Facility Registry Service}
\acro{fptz}[FPTZ]{Fastest Path to Zero Initiative}
\acro{stand}[STAND]{Siting Tool for Advanced Nuclear Development}
\acro{neup}[NEUP]{Nuclear Energy University Program}
\acro{um}[UM]{University of Michigan}
\end{acronym}    
\usepackage{amsmath,amsthm,amssymb,epsfig,bm,amsmath}
\usepackage{bm}
\usepackage{array}
\usepackage{graphicx}
\usepackage{cleveref}
\usepackage{graphicx}
\usepackage{float}
\usepackage{csquotes}
\usepackage{amsmath}
\usepackage{mhchem}
\usepackage{xparse}
\usepackage{adjustbox}
\usepackage{MnSymbol}
\usepackage{mathdots}
\usepackage{makecell}
\usepackage{listings}
\usepackage{footnote}
\usepackage[flushleft]{threeparttable}
\usepackage{multirow}
\usepackage{relsize}
\usepackage{lettrine}
\usepackage{courier}
\usepackage{comment}
\usepackage{soul} 
\sethlcolor{yellow} 
\usepackage{color} 
\definecolor{mygreen}{RGB}{28,172,0} 
\definecolor{mylilas}{RGB}{170,55,241}
\usepackage{amsmath,amsthm,amssymb,amsfonts} 
\usepackage{graphicx} 
\usepackage[margin=1in]{geometry} 
\usepackage[yyyymmdd]{datetime}
\usepackage{algorithm}
\usepackage[noend]{algpseudocode}
\usepackage{lipsum}
\usepackage{setspace}
\usepackage{siunitx}
\usepackage{pdfpages}
\usepackage{tikz}
\usepackage{hhline}
\usepackage{geometry}
\usepackage{longtable}
\usetikzlibrary{positioning}
\usepackage{booktabs,caption}
\usepackage[]{hyperref}
\hypersetup{
 colorlinks  = true, 
 urlcolor   = blue, 
 linkcolor  = black, 
 citecolor  = blue 
}
\usepackage[utf8]{inputenc}
 
\usepackage{listings}
\usepackage{color}
\definecolor{codegreen}{rgb}{0,0.6,0}
\definecolor{codegray}{rgb}{0.5,0.5,0.5}
\definecolor{codepurple}{rgb}{0.58,0,0.82}
\definecolor{backcolour}{rgb}{0.95,0.95,0.92}
\lstdefinestyle{mystyle}{
  backgroundcolor=\color{backcolour},  
  commentstyle=\color{codegreen},
  keywordstyle=\color{magenta},
  numberstyle=\tiny\color{codegray},
  stringstyle=\color{codepurple},
  basicstyle=\footnotesize,
  breakatwhitespace=false,     
  breaklines=true,         
  captionpos=b,          
  keepspaces=true,         
  numbers=left,          
  numbersep=5pt,         
  showspaces=false,        
  showstringspaces=false,
  showtabs=false,         
  tabsize=2,
  escapeinside={<@}{@>},
}
 
\usepackage{booktabs} 
\usepackage{siunitx} 

\usepackage{mathtools} 
\usepackage{gensymb} 
\usepackage{mhchem} 
\usepackage{fixltx2e} 
\usepackage{bbm}

\usepackage{multirow}

\usepackage{fancyhdr} 
\theoremstyle{definition}

\theoremstyle{definition}

\theoremstyle{remark}

\usepackage{soul} 
\soulregister\cite7
\soulregister\ref7
\soulregister\pageref7
\usepackage{bm}

\usepackage{framed} 

\usepackage{multicol} 

\usepackage{nomencl} 
\usepackage{tikz}
\makenomenclature
\usepackage[resetlabels,labeled]{multibib}

\renewcommand*\nompreamble{\begin{multicols}{2}}

\renewcommand*\nompostamble{\end{multicols}}

\usepackage{amssymb}
\usepackage{pifont}
\definecolor{light-gray}{gray}{0.95}

\makeatletter
\AtBeginDocument{\let\hl\@firstofone}
\makeatother

\usepackage{verbatim}

\journal{Energy Conversion and Management:X (published in \href{https://doi.org/10.1016/j.ecmx.2025.100923}{10.1016/j.ecmx.2025.100923})}

\begin{document}

\begin{frontmatter}

\title{\large Multi-objective Combinatorial Methodology for Nuclear Reactor Site Assessment: A Case Study for the United States}

\author{Omer Erdem$^{a}$, Kevin Daley$^{a}$, Gabrielle Hoelzle$^{a}$, Majdi I. Radaideh$^{a,*}$}

\cortext[mycorrespondingauthor]{Corresponding Author: Majdi I. Radaideh (radaideh@umich.edu)}

\address{$^{a}$Department of Nuclear Engineering and Radiological Sciences, University of Michigan, Ann Arbor, MI 48109, United States}

\begin{abstract}

As clean energy demand grows to meet sustainability and net-zero goals, nuclear energy emerges as a reliable option. However, high capital costs remain a challenge for nuclear power plants (NPP), where repurposing coal power plant sites (CPP) with existing infrastructure is one way to reduce these costs. Additionally, Brownfield sites—previously developed or underutilized lands often impacted by industrial activity—present another compelling alternative. This study introduces a novel multi-objective optimization methodology, leveraging combinatorial search to evaluate over 30,000 potential NPP sites in the United States. Our approach addresses gaps in the current practice of assigning pre-determined weights to each site attribute that could lead to bias in the ranking. Each site is assigned a performance-based score, derived from a detailed combinatorial analysis of its site attributes. The methodology generates a comprehensive database comprising site locations (inputs), attributes (outputs), site score (outputs), and the contribution of each attribute to the site score.  We then use this database to train a neural network model, enabling rapid predictions of nuclear siting suitability across any location in the United States. \hl{Our findings highlight that CPP sites are highly competitive for nuclear development, but some Brownfield sites are able to compete with them. Notably, four CPP sites in Ohio, North Carolina, and New Hampshire, and two Brownfield sites in Florida and California rank among the most promising locations.} These results underscore the potential of integrating machine learning and optimization techniques to transform nuclear siting, paving the way for a cost-effective and sustainable energy future.

\end{abstract}


\end{frontmatter}


\setstretch{1.2}

\section*{Highlights}

\begin{itemize}
    \item Analysis of the suitability of 30,000+ United States Brownfield and coal sites for hosting nuclear power plants.
    \item Multi-objective combinatorial methodology for nuclear power plant site ranking is developed.
    \item The method does not require analyst-defined weights in site objectives that could introduce bias.
    \item Data-driven neural network model is developed to predict site metrics for any location in the United States.
    \item \hl{Coal sites prove to be competitive for new nuclear plants, with some Brownfield sites also being feasible.} 
\end{itemize}

\section{Introduction}
\label{sec:intro}

Nuclear energy has several important applications, ranging from electricity generation to specialized industrial and medical uses \cite{orhan2012integrated,rosen2020nuclear,jarrah2019determination,pruavualie2018nuclear}. Nuclear energy ranks among the highest-yielding energy sources and has very low carbon dioxide emissions, even when accounting for the entire lifespan of the \ac{npp}. With its significant potential, nuclear energy could play a key role in achieving net-zero carbon emissions. However, a major challenge is its high levelized cost of electricity. While production costs for renewable energy sources have consistently declined over time, nuclear energy had the highest levelized cost of electricity in 2023, driven by rising capital expenditures and regulatory requirements \cite{Lazard}. In contrast, wind and solar energy tend to reduce electricity prices in their regions since their levelized costs are generally lower than the average for all energy sources \cite{Maciejowska}. Although the raw material costs for nuclear electricity production have also decreased \cite{Du}, the substantial initial costs of nuclear power facilities remain a primary factor in nuclear energy’s higher prices.

\hl{In addition to economic challenges, socioeconomic and regulatory factors significantly influence nuclear energy adoption. Public sentiment, political support, and regulatory hurdles shape perceptions of nuclear energy and its feasibility as a sustainable energy solution \cite{smolinski2024nuclear, hoedl2019social}. These factors also impact the repurposing of coal-fired power plants (CPPs) to Brownfield sites for nuclear energy projects, where public trust, local job creation, and community engagement play critical roles in project acceptance and success \cite{verma2024sociotechnical,kwon2024sentiment}. Without addressing these challenges through comprehensive policies and transparent communication, the adoption of nuclear energy and the redevelopment of CPP sites for nuclear projects may face considerable resistance \cite{smolinski2024nuclear}.} Furthermore, concerns around radioactive waste, safety, and infrastructure costs remain central to public discourse and influence both nuclear energy deployment and the feasibility of transforming CPPs into Brownfield nuclear sites \cite{kwon2024sentiment,price2019advanced}. \hl{In order to increase public acceptance of NPPs and ensure public cooperation, as well as to avoid public backlash, the site of the CPP transformed into an NPP should be selected carefully, with public sentiment toward nuclear energy taken into consideration.}

By repurposing established facilities, the capital costs associated with \ac{npp} construction could be reduced. This analysis focuses on the United States, where we consider two existing facility options for new nuclear siting: \ac{cpp}s and Brownfields. First, we consider using the \ac{cpp} station sites for \ac{npp} development. \hl{CPPs include control buildings, storage facilities, auxiliary structures, roads, electrical and water infrastructure, and substations. The presence of these existing facilities reduces the need for new construction, thereby lowering the capital costs associated with NPP installation.} In comparison to a greenfield \ac{npp} building operation, the overall capital costs could be decreased by 15-35\%, depending on the magnitude and choice of technology used in repurposing the framework of a coal power facility with a 1,200 MWe power output \cite{Hansen_inv}. However, these \ac{cpp} locations still have drawbacks. For example, if site remediation of a \ac{cpp} is required before beginning the \ac{npp} construction project, the coal to nuclear repurposing costs could be high. Still, the existing educated workforce of the \ac{cpp} would decrease the training and education costs of the personnel \cite{Hansen_gui}. A complete cost analysis must be carried out before such a repurposing operation. According to the U.S. Department of Energy, 237 operational and 157 idle potential CPP sites are appropriate for hosting nuclear energy facilities \cite{Hansen_inv}. Before reusing the \ac{cpp} buildings, each of these sites must be thoroughly studied, as \ac{npp}s are complex, long-term projects with several risk factors and consequences. The \ac{npp} sitting in U.S. \ac{cpp} locations has been researched in the past \cite{Erdem, Baskurt, Devanand, Locatelli}. CPP locations promise to be valuable sites for \ac{npp} siting. However, the number of U.S. \ac{cpp}s is not enough to achieve the ultimate net-zero goals for the country. Therefore, an analysis that includes a more comprehensive look at all potential \ac{npp} sites in the U.S. is needed.

The second option to host a NPP is to consider Brownfield sites. The term "Brownfield" refers to formerly built land that may be contaminated, underutilized, or abandoned, and may occasionally have low pollution levels \cite{Gray}. The economic costs and risks of Brownfield redevelopment projects are lower in areas that have an increased demand for residential and commercial development \cite{De_Sousa}. According to the U.S. \ac{epa}, \ac{acres}, Brownfield sites are properties where the existence or potential presence of a hazardous material, pollutant, or contaminant may make it more difficult to expand, redevelop, or reuse \cite{EPA}. Common types of Brownfield sites are former industrial or manufacturing plants, abandoned gas stations, landfills or waste disposal areas, rail yards, mines, and junkyards. Repurposing these sites could benefith both the environment and the economy. To assess, clean up, and sustainably reuse the Brownfield sites, the \ac{epa} Brownfields Program provides grants and technical assistance to communities.

The siting problem of \ac{npp} location has been an ongoing topic for the last 80 years. The problem consists of many different degrees of freedom and each requirement must be analyzed carefully since \ac{npp}s are long-term and high-cost projects. Most of the siting analyses in this field use a technique called the ``weighted sum method''. In this method, explicit weights are applied to the geographical data to perform location selection studies for \ac{npp}s. To form an objective function, these techniques use pre-determined weights that the analyst justifies based on several factors. Geographical characteristics like seismicity, the presence of cooling water, fault lines, the distance to centers of population, international borders, and logistic network are the most crucial factors taken into account in this weighted sum method \cite{Baskurt}. In some studies, only the most crucial characteristics are taken into account, such as seismic activity, population density, cooling water cost, reactor unit cost, and consumer proximity \cite{Devanand}. More sophisticated approaches use detailed \ac{npp} socioeconomic data. These methods handle financial, site-related, welfare, and project life-cycle factors independently in separate subroutines, by using their built-in socioeconomic data and weights \cite{Locatelli}. Both simple and complex approaches still depend on users to assess the weights assigned to each site objective or attribute. The analyst responsible for setting thresholds, assigning weights, and processing raw data ultimately makes these critical decisions. In this process, the choice of weights can be subjective or objective, but never entirely removes the analyst's bias in prioritizing certain sites for location selection. This is demonstrated in a recent study on \ac{cpp} assessment in the U.S. for \ac{npp} transition conducted by \cite{Rafi}, which involved assigning weights to various site objectives. While we recognize that the authors determined these weights through thorough research and extensive consultations, this level of rigor cannot be guaranteed in every study. Additionally, the approach still introduces bias, as site rankings inherently depend on the assigned weights. Therefore, an alternative method that avoids predetermined weights is needed to better identify ideal locations for \ac{npp} siting with limited bias. Multi-objective optimization techniques may offer some solution pathways to these challenges. 

A range of optimization methods for minimization and maximization problems are available in the literature, typically grouped into four main categories: stochastic, genetic, first-order, and second-order algorithms \cite{Kochenderfer}. Most optimization algorithms can solve single-objective optimization problems when the objective function (also known as fitness or cost function) is a scalar. However, when addressing problems that involve multiple objectives, \ac{moo} approaches are required. These algorithms are designed to handle the complexity of balancing multiple objectives simultaneously during the search.

\ac{ga}s are powerful optimization techniques that mimic natural selection, using processes such as mutation, crossover, and selection to evolve solutions to complex problems. In multi-objective \ac{ga}, there are two main approaches: elite-preserving and non-elite-preserving. Elite-preserving \ac{ga}s retain the best solutions—referred to as "elite" individuals—across generations, ensuring that high-quality solutions are not lost during the evolution process. This approach increases the likelihood of convergence toward optimal solutions by continuously preserving the strongest candidates. In contrast, non-elite-preserving GAs do not explicitly retain these top individuals, which can lead to a greater diversity in solutions but may slow convergence. Elite-preserving multi-objective GAs are often more effective in applications where maintaining high-quality solutions across generations is crucial. Non-elite-preserving methods are better suited to scenarios where exploration is prioritized. Multi-objective GAs like \ac{nsgaii} \cite{Deb_II} and \ac{nsgaiii} \cite{Deb_III} maintain the elites over iterations. The \ac{nsgaiii} is slightly superior to the \ac{nsgaii}. For instance, \ac{nsgaiii} uses the variable of crowding distance to identify additional population solutions and rank them in order to maintain variety.

In the problem of site selection, there is a high likelihood that multiple sites may have favorable characteristics which makes optimization challenging. We propose the concept of site characteristic ``domination'' as a solution to this problem. The method of non-dominated sorting is used in different optimization algorithms to identify optimal solutions in \cite{Deb_II, Deb_III, Kalyanmoy}. The most optimal samples in a non-dominated sorting operation are also known as the Pareto front, where each solution represents a trade-off across multiple objectives. In this Pareto-efficient set, no single solution is superior across all objectives, but the Pareto solutions that balance all objectives can be further analyzed. \hl{Consequently, in order to find the best NPP siting locations independent of any bias, we conduct a combinatorial analysis of Pareto front solutions by running a massive number of site attribute combinations, thereby mitigating external bias towards any site. This approach enables the identification of a ranked set of locations and the selection of the best site, rather than yielding inconclusive results in very high dimensional pareto fronts.}

This research presents a new approach to nuclear reactor site assessment and Brownfield redevelopment by specifically targeting the siting of nuclear power plants \ac{npp}s; an area that has not been thoroughly explored in the existing literature. Utilizing raw site data available through the Siting Tool for Advanced Nuclear Development (STAND) tool \cite{weir2023siting}, we created a dataset of Brownfield \ac{npp} siting locations, focusing on 34,211 sites that do not require extensive industrial cleanup, thereby establishing a robust resource for future nuclear siting projects. Furthermore, we conducted a comparative analysis between \ac{cpp}s and Brownfields, challenging the prevailing assumption that \ac{cpp} sites are the only sites that offer optimal conditions for \ac{npp} siting. By comparing 265 \ac{cpp}s with the Brownfield sites, our findings illuminate the feasibility of alternative sites, thereby offering additional compelling options to the current trends in coal-to-nuclear transition research. \hl{To enhance the efficiency of future site assessment, we developed a machine learning model that predicts siting metrics and objective importance; significantly reducing the computational burden associated with traditional methodologies.} This model is poised to facilitate rapid assessments for any contiguous U.S. location, ultimately streamlining the decision-making process in \ac{npp} siting. Accordingly, the key contributions of this study can be summarized as:
\begin{enumerate}
    \item A novel multi-objective combinatorial methodology is introduced for site ranking, eliminating the need for analyst-defined weights that could introduce bias. The approach is generic and adaptable, allowing it to be applied to other countries beyond the United States, where this study is based.
    
    \item This study represents the first comprehensive evaluation of such a large number of potential nuclear reactor sites—encompassing both Brownfield and coal sites—within the United States using a unified and flexible methodology.
    
    \item The extensive dataset and results generated through this search methodology have enabled the development of a machine learning model. This model facilitates rapid assessments of nuclear reactor site suitability across the United States, requiring only the site's coordinates, county, and state.
\end{enumerate}

The remaining sections are organized as follows: Section \ref{sec:data} delves into the data collection and processing, providing an overview of the socioeconomic, safety, and proximity site characteristics relevant to our study, along with a detailed examination of coal power plant sites. In Section \ref{sec:method}, we outline the research methodology, which encompasses a discussion of the Non-dominated Multi-objective Combinatorial Search and the implementation of Concatenated Neural Networks (ConcNN) to analyze the collected data. Following this, Section \ref{sec:results} presents the Results and Discussions, where we interpret the findings and their implications. Finally, Section \ref{sec:conc} concludes the paper, summarizing key insights and suggesting avenues for future research.

\section{Data Collection and Processing}
\label{sec:data}

\begin{figure}
    \centering
    \includegraphics[width=\textwidth]{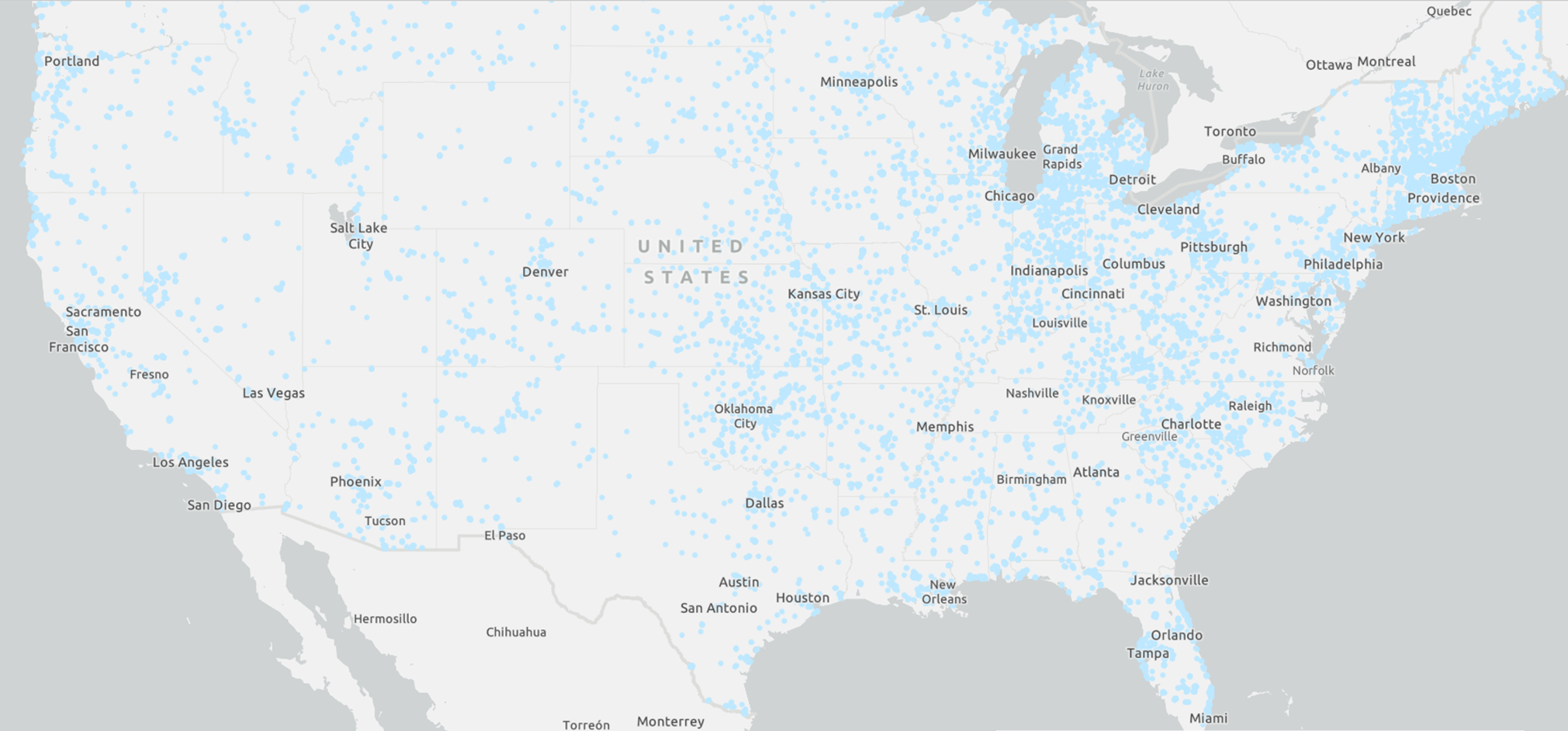}
    \caption{\ac{acres} Brownfield locations in the United States provided by the Environmental Protection Agency.}
    \label{fig:us_Brownfields_map}
\end{figure}

In the Brownfield nuclear reactor siting analysis, we use the \ac{acres} Brownfield database (described above) as a starting point. We gathered the geographical database files from the \ac{epa}. We extracted the coordinates recorded in \ac{acres} from the database layers. We then use the extracted coordinates in the \ac{stand} tool \cite{weir2023siting} developed by the National Reactor Innovation Center at Idaho National Laboratory and maintained by the Fastest Path to Zero Initiative at the University of Michigan \cite{FPTZ}. We processed each extracted Brownfield coordinate in STAND to creating a Brownfield nuclear reactor siting dataset that includes all Brownfield sites with their corresponding coordinates and site characteristics. We found unreliable data for Alaska and Hawaii in the \ac{stand} tool, because of data availability issues; the Brownfield sites in these two states were excluded from the final dataset. The \ac{acres} U.S. Brownfield locations gathered from \ac{epa} are given in Figure \ref{fig:us_Brownfields_map}. The Brownfield siting dataset is current as of May 2024. 

About 34,211 different contiguous U.S. Brownfield coordinates in the \ac{epa} \ac{acres} Brownfield database have been processed through STAND to generate values for 22 unique objectives. The attributes in the dataset describe the characteristics of each site. We chose these objectives to match the 1000 MWt or higher power reactor siting parameters used in the coal-to-nuclear transition analysis performed by the previous \ac{npp} siting research \cite{Rafi}. We classify the objectives selected for the Brownfield sites under three categories:

\begin{enumerate}
    \item \textit{Socioeconomic characteristics}: (1) Number of nuclear restrictions in the state, (2) state electricity price, (3) state net electricity imports, (4) state nuclear inclusive policy, (5) population sentiment towards nuclear energy, (6) traditional regulation in the energy market, (7) 5-year average labor rate, and (8) \ac{svi}.
    \item \textit{Safety characteristics}: (1) Number of intersecting protected lands, (2) number of hazardous facilities in 5 miles, (3) no fault line intersection, (4) no landslide area, (5) having a peak ground acceleration lower than 0.3g, (6) not having a flood in previous 100 years, (7) no open water or wetland intersection, and (8) having a slope lower than 12 percent.
    \item \textit{Proximity characteristics}: (1) Having a close population center, (2) having a close retiring power station, (3) having a close Research and Development (R\&D) center, (4) having a close electricity grid substation (5), transportation system distance, and (6) having a streamflow with 50 kgpm flow inside 20 miles.
\end{enumerate}

\subsection{Rationale for Objective Selection}

\hl{The STAND tool identifies 22 critical socioeconomic, safety, and proximity characteristics as key attributes for nuclear power plant (NPP) siting \cite{weir2023siting}. These characteristics, selected for their relevance, are justified in the Site Comparison step within the Relevance Form tab of the tool. The selection of these attributes reflects a comprehensive understanding of technical feasibility, safety imperatives, and socioeconomic considerations, ensuring that the siting process aligns with both regulatory requirements and public expectations. By integrating these diverse criteria, the STAND tool facilitates informed decision-making for sustainable and publicly acceptable nuclear energy development \cite{iaea2022managing, basu2019site}.

Several \textbf{socioeconomic} attributes significantly influence NPP siting. State-level restrictions present a key challenge, ranging from procedural barriers to outright moratoria, requiring substantial time and resources to address \cite{basu2019site}. Retail electricity prices act as a critical indicator of nuclear energy's cost competitiveness in specific states \cite{schuelke2014socio}. States with energy consumption exceeding their production are more inclined to prioritize new energy facilities, including NPPs, to ensure energy security \cite{schuelke2014socio}. Public attitude toward nuclear energy is essential for community engagement and the success of siting projects \cite{shields1979socioeconomic}. Additionally, market regulation affects factors such as capital costs, financing, and industry partnerships, which shape the feasibility of NPP siting \cite{iaea2022managing}. Local labor rates further influence the economics of construction and maintenance \cite{schuelke2014socio}. Finally, the inclusion of the Social Vulnerability Index (SVI) ensures equitable site selection by addressing risks to vulnerable populations and promoting resilience \cite{frantal2016distance}.

\textbf{Safety} is paramount in NPP siting, requiring thorough evaluations of hazardous facilities, seismic risks, and other geotechnical factors. Proximity to industrial sites, such as chemical plants or refineries, introduces risks including toxic releases or flammable materials, necessitating careful assessment \cite{basu2019site}. Seismic considerations, such as fault lines within a 200-mile radius and ground motion potential, significantly impact site suitability and design requirements \cite{10CFR100AppendixA}. Site-specific factors like liquefaction, landslides, and subsidence are critical for ensuring the integrity of safety-related functions \cite{iaea2022managing}. Wetlands and floodplains are evaluated under regulations like Executive Order 11988 and the Clean Water Act to avoid ecological damage and mitigate flooding risks \cite{basu2019site}. Additionally, steep slopes are avoided due to the high economic costs and technical challenges associated with construction \cite{NRC_ML23123A090}.

\textbf{Proximity} considerations ensure NPPs are located in areas conducive to operational efficiency and minimal risk. Regulatory guidelines, such as U.S. Nuclear Regulatory Commission (NRC) Regulatory Guide 4.7, recommend population density thresholds to ensure emergency preparedness and safety \cite{10CFR100AppendixA}. Reusing infrastructure from closed facilities can streamline development and reduce costs \cite{basu2019site}. Proximity to nuclear research facilities enhances technical support, while access to electrical substations ensures seamless integration with the power grid \cite{iaea2022managing}. Transportation infrastructure, including heavy-haul routes, is critical for delivering modular reactor components \cite{basu2019site}. Finally, freshwater availability is essential for reactor cooling, making proximity to adequate water resources a key criterion for site selection \cite{basu2019site}.

The STAND tool provides a holistic framework for NPP siting, by addressing these socioeconomic, safety, and proximity factors. This approach balances technical feasibility with safety imperatives and socioeconomic considerations, fostering public acceptance and compliance with regulatory requirements. The integration of these criteria into our analysis enhances the alignment of nuclear energy development with sustainability goals and community resilience, thereby strengthening the role of nuclear power in the transition to clean energy \cite{iaea2022managing, basu2019site, schuelke2014socio, 10CFR100AppendixA}.}

\subsection{Socioeconomic Characteristics}

We extracted nuclear restrictions in each state. Considering it is an uncontrollable parameter for a stakeholder, we define a moratorium as a deal breaker. We express all other restrictions in a single variable as the number of nuclear restrictions in the state. These five remaining restrictions are state legislature approval, state commissioner of environmental protection approval, proving that the construction of a nuclear facility will be economically feasible for ratepayers, demonstrable technology or a means for high-level waste disposal or reprocessing, and proving that the radioactive waste material disposal method will be safe. 

We define state electricity price as the annual average electricity price in cents/kWh and electricity imports as the annual amount of electrical power imported by the state in millions of kWh. Our ``population sentiment towards nuclear energy" parameter is an average of the 10-year poll findings of sentiment values of the counties in any 20-mile radius. We set state nuclear inclusive policy as a binary value that shows if the state administration has any decisions to support the development of nuclear energy, such as Renewable Portfolio Standards (RPSs), Renewable Portfolio Goals (RPGs), and Clean Energy Standards (CESs). 

In energy market regulation, traditional regulation is considered a positive attribute since a regulation authority would help nuclear energy to become the stable baseline in a grid. The construction labor rate is the 5-year average of state labor rate, provided by Occupational Employment and Wage Statistics, in USD.

\subsection{Safety Characteristics}
We selected safety objectives as characteristics that support siting a 1000MWt or higher power reactor \cite{Rafi}. Most of the safety objectives (1, 3, 4, 6, 7) are intersections with dangerous areas explained in the list of objectives provided earlier in this section. NRC Regulatory Guide 4.7, ``General Site Suitability Criteria for Nuclear Power Stations'' describes that siting nuclear reactors close to the publicly used lands depends on local jurisdictions and can be refused. For this reason, the protected land objective checks if the site intersects with American Indian reservations, correctional facilities, critical habitats, forests, hospitals, national monuments, national or state parks, schools or colleges, wild and scenic rivers, wilderness areas, or wildlife refuges. If the site intersects with any of these, the number of intersections is given. 

The objective of ``having a near hazardous facility within a 5-mile radius'' checks for any petroleum or gas processing facility, pumping stations, fertilizer plants, chemical plants, nuclear fuel plants, storage tanks, and airports. For the fault lines, 10 CFR 100, Appendix A, Table 1 specifies that fault lines within 200 miles of the site must be considered \cite{10CFR100AppendixA}. The STAND tool checks if the location is within 200 miles of any fault line. The United States Geological Survey (U.S.GS) reports the lands that have a moderate or high risk of landslides. This objective states whether the location intersects with a high or moderate landslide-risk area, if any. The peak ground acceleration objective checks for any ground acceleration history higher than 0.3g in the area, which is the suggested limit for the high-power light water reactor (LWR) installations in the 2002 EPRI (Electric Power Research Institute) siting guidance \cite{EPRI2002Siting}. 

The flood zone objective checks if the site lies in a 100-year floodplain. For the open waters and wetlands, depending on the region of the open water, these areas are protected by (1) the Rivers and Harbors Act, (2) the Wild and Scenic Rivers Act, (3) the Clean Water Act, (4) the Coastal Management Act, (5) the Coastal Barriers Resources Act, (6) the Fish and Wildlife Coordination Act, (7) the Migratory Bird Conservation Act, and (8) the National Wildlife Refuge Act. This objective checks if the site intersects with lakes, rivers, streams, ponds, etc. The objective featuring the slope of the site is a binary variable that returns 1 if the slope of the site location is higher than 12\% as recommended in the 2002 EPRI siting guidance \cite{EPRI2002Siting}. 

\subsection{Proximity Characteristics}
Proximity objectives have been used as binary values as suggested by the STAND tool and previous research conducted in the area. For example, the 10 CFR 100 Regulatory Guide of the U.S. NRC shows that at least a four-mile distance is required between the \ac{npp} and the population centers with 25,000 residents \cite{Belles, NRC10CFR100Guide}. In the STAND tool, this objective is set to 0 if the closest population center distance is closer than four miles. For centers with more than four miles, the objective is linearly scaled between [0,1], where the center with the highest distance from the plant is set to 1.

Communities with existing nuclear technology are generally more supportive of nuclear technology as \ac{npp}s are an important source of local jobs and tax revenues. Operating nuclear facilities show the number of nuclear facilities or generators within a 100-mile radius. Having access to nearby nuclear R\&D facilities can provide technical support during the development of a nuclear facility. The R\&D center objective shows if there is a nuclear R\&D center within 100 miles. 

The retiring power station objective shows if any coal, natural gas, or nuclear power station is in a one-mile radius in the next 20 years. For the sites with a retiring power station nearby, the distances are linearly scaled to the [0,1] range where the lowest-distance site is set to 1. The substation distance objective is the site distance to the closest electricity grid substation in miles, scaled to [0,1] range where the closest site to the substation is set to 1. We define transportation systems as major roads in the United States. The objective is set to 1 if there is a major road closer than 1 mile. For the sites that do not have a major road in a 1-mile radius, the values are linearly scaled to the [0,1] range where the lowest-distance site is set to 1. The streamflow objective shows if there are any freshwater supplies in 20-mile radius, with the minimum 50,000 gallons per minute (GPM) streamflow as recommended in the STAND tool. 

The distance values in this section are the distance values that the STAND tool uses for thresholding and linearly scaling the attributes of the selected sites. We selected the generator retirement time (20 years), transportation method (major roads), and streamflow GPM value (50,000) with the guidance of the STAND tool developers. 

Lastly, an example of the created dataset is given in Figure \ref{tab:data_example}. This dataset is scaled to the [0,1] range afterwards, higher values for each attribute indicate favorable objectives. For example, 0.9 in the state electricity imports column indicates a high electricity import in the state, while 0.1 indicates low electricity import.

\begin{table}[!h]
\centering
\caption{Preprocessed data with three examples for Brownfield sites.}
\begin{tabular}{|l|c|c|c|}
\hline
\textbf{Attribute} & \textbf{Site A} & \textbf{Site B} & \textbf{Site C} \\ \hline
Registry ID & 110000339982 & 110000344832 & 110000346028 \\ \hline
Longitude & -76.56465 & -81.45892 & -80.25551 \\ \hline
Latitude & 39.27712 & 39.40018 & 35.82021 \\ \hline
County FIPS & 24510 & 54107 & 37057 \\ \hline
State FIPS & 24 & 54 & 37 \\ \hline
Number of Nuclear Restrictions in the State & 0 & 0 & 0 \\ \hline
State Electricity Price & 14.30538462 & 10.26 & 10.62692308 \\ \hline
State Net Electricity Imports & 25934 & -28050 & 14875 \\ \hline
State Nuclear Inclusive Policy & 0 & 0 & 0 \\ \hline
Population Sentiment Towards Nuclear Energy & 0.402230895 & 0.428866209 & 0.658312001 \\ \hline
Traditional Regulation In The Energy Market & 1 & 1 & 1 \\ \hline
5-Year Average Labor Rate & 34766 & 35606 & 30384 \\ \hline
Social Vulnerability Index & 0.41585454 & 0.38254449 & 0.433715808 \\ \hline
Number of Intersecting Protected Lands & 0 & 1 & 0 \\ \hline
Number of Hazardous Facilities in 5 Miles & 8 & 6 & 3 \\ \hline
No Fault Line Intersection & 1 & 1 & 1 \\ \hline
No Landslide Area & 0 & 0 & 0 \\ \hline
Having A Peak Ground Acceleration Lower Than 0.3g & 1 & 1 & 1 \\ \hline
Not Having A Flood in Previous 100 Years & 0 & 0 & 1 \\ \hline
No Open Water Or Wetland Intersection & 1 & 1 & 1 \\ \hline
Having A Slope Lower Than 12 Percent & 1 & 1 & 1 \\ \hline
Population Center Distance & 3.204225645 & 13.69500274 & 13.22706011 \\ \hline
Retiring Facility Distance & 2.951268247 & 76.8295757 & 52.97458529 \\ \hline
Existing Nuclear R\&D Center in 100 Miles & 1 & 0 & 1 \\ \hline
Electricity Substation Distance & 1.381065329 & 1.642955799 & 1.266471271 \\ \hline
Transportation System Distance & 0.778196965 & 1.49198691 & 1.235386683 \\ \hline
Having A Streamflow with 50 Kgpm Flow in 20 Miles & 1 & 1 & 1 \\ \hline
\end{tabular}
\label{tab:data_example}
\end{table}

\subsection{Coal Power Plant Sites}
According to the U.S. Department of Energy, there are currently 157 inactive and 237 active candidate coal power plant locations that are suitable for housing \ac{npp}s \cite{Hansen_inv}. These sites are accompanied by pertinent data, including the 22 objectives outlined in the \ac{stand} framework, which are detailed in previous sections and in Table \ref{tab:data_example}. 

Previous research identified 265 U.S. coal power plant sites as relevant for \ac{npp} siting \cite{Rafi}. Additionally, the coal-to-nuclear transition dataset was updated in October 2024 by the Fastest Path to Zero Initiative at the University of Michigan. We have adapted the coal power plant dataset, which was previously examined using the weighted sum method, for multi-objective optimization. By using this dataset, we facilitated a one-to-one comparative analysis with the Brownfield sites.

\hl{The coal-to-nuclear siting dataset is considerably smaller than the Brownfield siting dataset, therefore the addition of the CPP dataset to the Brownfield dataset does not significantly increase the processing time. We have created a joint dataset with Brownfields and CPPs to analyze and compare them. The joint dataset has then been used to train the machine learning models described later in the next section. Processing all sites as a single dataset allowed a comparison between the optimal locations identified in both datasets.}

\section{Research Methodology}
\label{sec:method}

This section describes the methodology used for the assessment of the Brownfield and coal sites described in Section \ref{sec:data}, which consists of two major parts. First, we describe our high-dimensional non-dominated combinatorial search approach that is used to rank all sites based on how many times they show up in the Pareto-front when considering every possible combination of the 22 site objectives. This search leads to a set of metrics that describe the importance of each objective to the rank of the site and a site score for global ranking. \hl{The methodology is applied to the joint dataset of 34211 Brownfield sites and 265 CPP sites.} Given our relatively large dataset of more than +30,000 sites, in the second part, we train a data-driven methodology using Concatenated Neural Networks (ConcNN) and lookup tables that allow predicting the (1) site score, (2) site objectives, and (3) importance value of each objective to the site score. The network needs only the site location to make such a prediction. 

\hl{The proposed methodology in this section provides a comprehensive, unbiased framework for aggregating results across all possible combinations of objectives, avoiding analyst-defined weights that might introduce subjective bias. This exhaustive approach ensures that the siting metric reflects a diverse range of objective trade-offs, which is particularly useful in high-dimensional decision-making problems. By leveraging non-dominated sorting for each subset of objectives, the method captures multi-objective interactions that might otherwise be overlooked, enhancing the robustness of the analysis.

The siting metric described in section \ref{sec:comb} provides a straightforward assessment of each site's performance relative to others in the dataset. Essentially, it functions as a dominance metric, quantifying the extent to which a site's objectives outperform those of other sites. This metric is purely mathematical, relying on the non-dominated sorting method, a widely used approach in multi-objective optimization problems.}

\subsection{Non-Dominated, Multi-Objective Combinatorial Search}
\label{sec:comb}

\hl{We used a combinatorial search between the 22 objectives to ensure a bias-free outcome through the application of a non-dominated sorting method. When we perform non-dominated sorting across all objectives at once, the resulting Pareto front comprises 1,294 locations, which yields inconclusive results for a single optimal site. To address this limitation, we applied non-dominated sorting to all possible combinations of objectives. A detailed explanation of these combinations is provided in Table \ref{tab:combs_explanation}, which outlines the combinations of objectives for each specified combination length.}

\begin{table}[ht]
\caption{Explanation of combination lengths and total number of combinations.}
\label{tab:combs_explanation}
\centering
\begin{tabular}{|c|l|c|}
\hline
\textbf{Combination Length ($s$)} & \textbf{Combination Examples ($L_{s,k}$)} & \textbf{Total Number (N)} \\ \hline
1  & (1), (2), (3), $\dots$, (21), (22) & 22 \\ \hline
2  & (1,2), (1,3), (1,4), $\dots$, (21,20), (21,22) & 231 \\ \hline
3  & (1,2,3), (1,2,4), $\dots$, (20,21,19), (20,21,22) & 1,540 \\ \hline
4  & (1,2,3,4), (1,2,3,5), $\dots$, (19,20,21,22) & 7315 \\ \hline
5  & (1,2,3,4,5), (1,2,3,4,6), $\dots$, (18,19,20,21,22) & 26,334 \\ \hline
6  & (1,2, $\dots$, 5,6), $\dots$, (17,18, $\dots$, 21,22) & 74,613 \\ \hline
7  & (1,2, $\dots$, 6,7), $\dots$, (16,17, $\dots$, 21,22) & 170,544 \\ \hline
8  & (1,2, $\dots$, 7,8), $\dots$, (15,16, $\dots$, 21,22) & 319,770 \\ \hline
9  & (1,2, $\dots$, 8,9), $\dots$, (14,15, $\dots$, 21,22) & 497,420 \\ \hline
10 & (1,2, $\dots$, 9,10), $\dots$, (13,14, $\dots$, 21,22) & 646,646 \\ \hline
11 & (1,2, $\dots$, 10,11), $\dots$, (12,13, $\dots$, 21,22) & 705,905 \\ \hline
12 & (1,2, $\dots$, 11,12), $\dots$, (11,12, $\dots$, 21,22) & 646,646 \\ \hline
13 & (1,2, $\dots$, 12,13), $\dots$, (10,11, $\dots$, 21,22) & 497,420 \\ \hline
14 & (1,2, $\dots$, 13,14), $\dots$, (9,10, $\dots$, 21,22) & 319,770 \\ \hline
15 & (1,2, $\dots$, 14,15), $\dots$, (8,9, $\dots$, 21,22) & 170,544 \\ \hline
16 & (1,2, $\dots$, 15,16), $\dots$, (7,8, $\dots$, 21,22) & 74,613 \\ \hline
17 & (1,2, $\dots$, 16,17), $\dots$, (6,7, $\dots$, 21,22) & 26,334 \\ \hline
18 & (1,2, $\dots$, 17,18), $\dots$, (5,6, $\dots$, 21,22) & 7,315 \\ \hline
19 & (1,2, $\dots$, 18,19), $\dots$, (4,5, $\dots$, 21,22) & 1,540 \\ \hline
20 & (1,2, $\dots$, 19,20), $\dots$, (3,4, $\dots$, 21,22) & 231 \\ \hline
21 & (1,2, $\dots$, 20,21), $\dots$, (2,3, $\dots$, 21,22) & 22 \\ \hline
22 & (1, 2, 3, $\dots$, 20, 21, 22) & 1 \\ \hline
Total Combination Sum: & & 4,194,303 \\ \hline
\end{tabular}
\end{table}

The preprocessed \hl{joint dataset (both Brownfield and coal sites)} contains 16,057 rows that represent the individual sites and 22 columns corresponding to objectives. The contiguous U.S. \ac{acres} Brownfield dataset includes sites that are often adjacent or in close proximity to one another. The data in the \ac{stand} tool rely on state-level and county-level information, proximity to other sites of interest, and integration with geographical databases. We expect locations within close proximity to exhibit identical characteristics. To manage this redundancy, we truncated site coordinates to 0.01 degrees of longitude and latitude, and we removed sites with identical truncated coordinates from the dataset. Later, we inferred the siting results for these removed locations by using results from persisting locations with the same truncated coordinates. This method introduces an uncertainty of approximately 0.315 miles in the southern U.S. and 0.246 miles in the northern U.S. due to the truncation of coordinates. \hl{The computational complexity of the non-dominated sorting method varies between $O(N \log N)$ and $O(N^2)$, depending on the implementation and problem structure. Reducing the number of total sites from 34,476 to 16,057 results in a 64.23\% decrease in computation time under the low-complexity $O(N \log N)$ scenario and a 77.85\% decrease under the high-complexity $O(N^2)$ scenario.}

The non-dominated, multi-objective combinatorial search starts from combination length \(s=1\). We only apply non-dominated sorting to a single selected objective at a time. For the combination length $s=1$, we apply non-dominated sorting  $k \in [1,N=22]$ times for the examples given in Table \ref{tab:combs_explanation}. \hl{For the combination length $s=2$, we apply non-dominated sorting  $k \in [1,N=231]$ times, and so on.}

The preprocessed dataset consists of \textbf{16,057} rows, represented as the matrix $\widehat{A} \in \mathbb{R}^{16057 \times 22}$. Subsets of $\widehat{A}$ that include only selected column indices are denoted as $\widehat{A}_{L_{s,k}}$, also referred to as $\widehat{B}_{s,k}$. Each $\widehat{B}_{s,k}$ matrix corresponds to a specific combination of selected columns, indexed by $s$ and $k$. To encapsulate all possible subsets, we construct a single tensor $\widehat{B}$ that consolidates this subset information.

We generated the possible combinations of the siting objectives using the combinatorial function $C(t,s)$. For $C(22,1)$, there are 22 different $L_{s,k}$ elements to select, each of these elements is given in Table \ref{tab:combs_explanation}. This operation creates $L$, the list of possible combinations that choose $s$ elements from a set of $t$ elements regardless of order. \hl{$L$ is a list of ordered index sets.} $L_{s,k}$ defines a single combination in the list $L$, where $L_{1,1}=(1)$, $L_{1,2}=(2)$, $L_{2,1}=(1,2)$ which is shown as:

\begin{equation}
\begin{split}
C(22,s) = L \\
L = \{L_{s,1}, L_{s,2}, L_{s,3}, \ldots, L_{s,N}\} \\
\text{where N is the total number of combinations for s.} \\
\end{split}
\end{equation}

The function $NS$ is defined as the operation that applies non-dominated sorting on a matrix or tensor. In Eq.\eqref{eq:NS_func}, $i$ represents the site index and $j$ represents the objectives of the selected subset ($\widehat{B}_{s,k}$). \hl{The NS function is applied along the site index ($i$) axis for each subset defined by the ($s,k$) pair.} The $NS$ function outputs a tensor $\widehat{r}$ with \hl{16,057 length in $i$ axis} indicating the domination for each site in the selected subset. \hl{For a given subset $\widehat{B}_{s,k}$}, $r_{s,k,i}$ is 1 for the ``dominating'' $\widehat{B}$ site indices ($i$) and 0 for the ``dominated'' site indices. For each matrix slice of $\widehat{B}$ corresponding to a given $(s, k)$ pair, there exists a unique vector slice of the $\widehat{r}$ tensor that represents the results associated with that specific $(s, k)$ pair.

\begin{equation}
\begin{aligned}
\widehat{B}_{s,k} = \widehat{A}_{L_{s,k}} \quad , \quad \widehat{B} \in \mathbb{R}^{22 \times N \times 16057 \times 22} \quad , \quad \text{with indices } (s,k,i,j) \\
NS(\widehat{B}) = \widehat{r} \quad , \quad \widehat{r} \in \mathbb{R}^{22 \times N \times 16057} \quad , \quad \text{with indices } (s,k,i).
\label{eq:NS_func}
\end{aligned}
\end{equation}

The next step involves the normalization of the tensor $\widehat{r}$ on its $k$ axis. Normalized $\widehat{r}$ tensor is shown as $\widehat{r'}$. The sum of $\widehat{r'}$ along the $k$ axis produces the normalized observation ratio matrix $\widehat{NR}$. 

\begin{equation}
\begin{split}
r'_{s,k,i} = \frac{r_{s,k,i}}{\sum_{k} r_{s,k,i}} \quad , \quad \forall s, i \\\widehat{r'} \in \mathbb{R}^{22 \times N \times 16057} \quad , \quad \text{with indices } (s,k,i) \\
NR_{s,i} = \sum_{k} r'_{s,k,i} \quad , \quad 
\widehat{NR} \in \mathbb{R}^{22 \times 16057} \quad , \quad \text{with indices } (s,i) .\\ 
\end{split}
\end{equation}

The final summed normalized observation ratio $\overrightarrow{SR}$ is obtained by summing $\widehat{NR}$ over all combination lengths $s$. Next, we apply min-max scaling to $\overrightarrow{SR}$ within the range $[0,1]$, which produces $\overrightarrow{\mathbb{M}}$ that acts as a site performance metric for selecting the most suitable reactor sites. These two variables are defined as follows:

\begin{equation}
\begin{split}
SR_i = \sum_{s=1}^{22} NR_{s,i} \quad , \quad \overrightarrow{SR} \in \mathbb{R}^{16057} \quad , \quad  \text{with index } (i)\\
\overrightarrow{\mathbb{M}} = \frac{\overrightarrow{SR} - \min(\overrightarrow{SR})}{\max(\overrightarrow{SR}) - \min(\overrightarrow{SR})}  \quad , \quad 
\overrightarrow{\mathbb{M}} \in \mathbb{R}^{16057} \quad , \quad  \text{with index } (i).
\end{split}
\end{equation}

The methodology presented is based on the non-dominated sorting approach, which we employed to identify the optimal individuals within a given set. Consequently, the siting metric derived from this method is inherently influenced by the specific sites included in the dataset. We demonstrate an example of acquiring the site observation ratio below. In this example, we select the 3-length combination of the objectives. 

\begin{equation*}
\begin{aligned}
 & C(22,3) = L[s=3,k=230] = [L_{3,1}, L_{3,2}, L_{3,3}, \ldots, L_{3,1540}] \\[1em]
&\text{Example:} \ L[s=3,k=230]  = \begin{bmatrix} 2, 3, 5 \end{bmatrix} \\[1em]
\end{aligned}
\end{equation*}

The function $C(s,t)$ creates the list of the combinations. We arbitrarily select the $230^{th}$ item on this list for demonstration. This list, $L_{s,k}[s=3,k=230]$, includes the column (objective) numbers 2, 3, and 5, which represent state electricity price, state electricity imports, and population \ac{svi}, respectively. The subset $\widehat{A}_{L_{s,k}}[s=3,k=230]$ of the dataset $\widehat{A}$ (or tensor $\hat{B}$ as defined before) is shown below:

\begin{equation*}
\begin{aligned}
\widehat{B}[s=3,k=230] & = \begin{bmatrix} 
0.2023 & 0.6605 & 0.6158\\
0.0729 & 0.2907 & 0.5883 \\
0.3935 & 0.5107 & 0.9067 \\
0.4777 & 0.7118 & 0.6409 \\
\vdots & \vdots & \vdots 
\end{bmatrix} \\[1em]
\end{aligned}
\end{equation*}

The subset $\widehat{B}[s=3,k=230]$ of $\widehat{B}$ shows a single matrix slice in the 4D $\widehat{B}$ tensor. Applying non-dominated sorting to the dataset creates $\widehat{r}$ tensor. Fixing the $s$ and $k$ variables and leaving the site index ($i$) variable free in this 3D tensor results in its vector slice $\widehat{r}[s=3,k=230]$:

\begin{equation*}
\begin{aligned}
NS(\widehat{B}[s=3,k=230]) & = \widehat{r}[s=3,k=230] = \begin{bmatrix} 
0 \\
0 \\
1 \\
1 \\
\vdots
\end{bmatrix} \\[1em]
\end{aligned}
\end{equation*}

This vector indicates that the sites at the first and second row indices are dominated by the other sites in this example. After finding the domination status of all locations for all 3-length combinations of the siting objectives (from 1 to 1,540 total combinations), we normalize the summed vector and sum for all k values. These operations create $\widehat{NR}$. Fixing the $s=3$ and $k$ variables and leaving the site index ($i$) variable free creates the vector $\widehat{NR}[s=3]$:

\begin{equation*}
\begin{aligned}
\widehat{NR}[s=3] & = \begin{bmatrix} 
0 \\
0 \\
0.0002 \\
0.0002 \\
\vdots
\end{bmatrix} \\[1em]
\end{aligned}
\end{equation*}

Summing $\widehat{NR}$ for all combination lengths ($s$) creates the summed normalized observation ratio $\overrightarrow{SR}$, which can then be converted by min-max normalization to $\overrightarrow{\mathbb{M}}$ (the siting metric) as follows:

\begin{equation*}
\begin{aligned}
\widehat{SR} & = \begin{bmatrix} 
0.013 \\
0.008 \\
0.029 \\
0.034 \\
\vdots
\end{bmatrix}
\end{aligned}
\end{equation*}

\begin{equation*}
\begin{aligned}
\overrightarrow{\mathbb{M}} & = \begin{bmatrix} 
0.0065 \\
0.0040 \\
0.0145 \\
0.0170 \\
\vdots
\end{bmatrix}
\end{aligned}
\end{equation*}

Following the site score, $\overrightarrow{\mathbb{M}}$, we seek to determine which of the 22 objectives contribute the most to the site score. This provides a more informative explanation for a given site's suitability. This information is shown in the summed normalized objective contributions matrix ($\widehat{SC}$), which shows the contribution of each objective (j) to a specific site performance (i). To gather this, we start by defining a slice of the dominating indices tensor $\widehat{r}$ from the previous derivation. This slice is named $\widehat{r}_{s,k}$ as a column vector. $\widehat{r}_{s,k}$ is multiplied with $\overrightarrow{v}_{s,k}^T$ row vector. The row vector $\overrightarrow{v}_{s,k}^T$ shows which objectives are included in the current combination: 

\begin{equation}
\begin{split}
\overrightarrow{v}_{s,k}[j] = 
\begin{cases} 
1 & \text{if } j \in L_{s,k} \\ 
0 & \text{if } j \notin L_{s,k} 
\end{cases}  \quad , \quad v_{s,k} \in \mathbb{R}^{22 \times 1}  \quad , \quad r_{s,k} \in \mathbb{R}^{16057 \times 1} \\
\widehat{R}_{s,k} = \widehat{r}_{s,k} \cdot \overrightarrow{v}_{s,k}^T \quad , \quad \widehat{R} \in \mathbb{R}^{22 \times N \times 16057 \times 22}  \quad , \quad \text{with indices } (s,k,i,j).
\end{split}
\end{equation}

The vector $\overrightarrow{v}_{s,k}^T$ has 1 at the indices included in the objective combination $L_{s,k}$ for the combination $k$. The 4D tensor $\widehat{R}$ results from merging the matrices for different $s$ and $k$ sets for each dot product result. $\widehat{R}$ shows which objectives ($j$) affected the performance of the sites ($i$). We define $\widehat{R}$ as the contribution matrix for $s^{th}$ combination length and the $k^{th}$ objective combination in the dataset. These contribution matrices are summed for different combinations within the same combination length as follows:

\begin{equation}
OC_{s,i,j} = \sum_{k} R_{s,k,i,j} \quad , \quad \widehat{OC} \in \mathbb{R}^{22 \times 16057 \times 22}.
\end{equation}

The objective contribution ($\widehat{OC}$) is row-normalized to not favor any combination length. The indices \(i\) and \(j\) represent the row and column indices of the matrix, respectively, and column indices ($j$) are normalized such that the sum of each row equals 1.

\begin{equation}
\widehat{NC}_{s,i,j} = \frac{\widehat{OC}_{s,i,j}}{\sum_{j} \widehat{OC}_{s,i,j}} \quad , \quad \widehat{NC} \in \mathbb{R}^{22 \times 16057 \times 22}.
\end{equation}

The normalized objective contributions ($\widehat{NC}$) are then summed for all combination lengths ($s$) to get the summed objective contribution ($\widehat{S}$). 

\begin{equation}
\begin{split}
\widehat{S}_{i,j} = \sum_{s=1}^{22} \widehat{NC}_{s,i,j} \quad , \quad \widehat{S} \in \mathbb{R}^{16057 \times 22}.\\
\end{split}
\end{equation}

The summed objective contribution ($\widehat{S}$) matrix is row-normalized to set the sum of importances to 1 for every site individually in the dataset. The aforementioned operations form the normalized summed objective contribution ($\widehat{SC}$). 

\begin{equation}
\begin{split}
\widehat{SC}_{i,j} = \frac{\widehat{S}_{i,j}}{\sum_{j} \widehat{S}_{i,j}}  \quad , \quad \widehat{SC} \in \mathbb{R}^{16057 \times 22}.
\end{split}
\end{equation}

An example of achieving the normalized objective contributions is given below. For the selected same objective combination ($k=230$) with the column numbers 2, 3 and 5, the non-dominated sorting result is given as \(\widehat{r}[s=3,k=230]\). Each row shows a site index which dominates the result of the non-dominated sorting.

\begin{equation*} 
\begin{split}
\widehat{r}[s=3,k=230] & = \begin{bmatrix}
0 \\
0 \\
1 \\
1 \\
\vdots
\end{bmatrix} \\[1em]
\end{split}
\end{equation*}

The combination selected for this example includes the columns 2, 3 and 5 of the dataset. The index set inside $L$ including these columns is given as \(L[s=3,k=230]\).

\begin{equation*} 
\begin{split}
L_{3,230} & = \begin{bmatrix}
2, 3, 5
\end{bmatrix} \\[1em]
\end{split}
\end{equation*}

We use the combination of objectives in this example to create a vector where an entry is 1 if the corresponding objective index is included and 0 otherwise. Note that the length of the vector \(\overrightarrow{v}[s=3,k=230]\) is 22, but only the indices 2, 3, and 5 are non-zero.

\begin{equation*} 
\begin{split}
\widehat{v}[s=3,k=230] & = \begin{bmatrix}
0 \\
1 \\
1 \\
0 \\
1 \\
\vdots
\end{bmatrix} \\[1em]
\end{split}
\end{equation*}

The matrix \(\widehat{R}_{s,k}\) is created from the dot product of \(\widehat{r}[s=3,k=230]\) and \(\widehat{v}[s=3,k=230]^T\).

\begin{equation*} 
\begin{split}
\widehat{R}[s=3,k=230] & = \begin{bmatrix}
0 & 0 & 0 & 0 & 0 & \hdots \\
0 & 0 & 0 & 0 & 0 & \hdots \\
0 & 1 & 0 & 1 & 1 & \hdots \\
0 & 1 & 0 & 1 & 1 & \hdots \\
\vdots & \vdots & \vdots & \vdots & \vdots
\end{bmatrix} \\[1em]
\end{split}
\end{equation*}

The contribution matrices of combinations \(\widehat{R}\) are summed for all existing combinations of the given combination length (e.g., \(s=3\)).

\begin{equation*} 
\begin{split}
\widehat{OC}[s=3] & = \begin{bmatrix}
0.3 & 0.5 & 2 & 0.7 & 0.1 & \hdots \\
5 & 1.1 & 0.6 & 0.1 & 0.3 & \hdots \\
0.2 & 1.2 & 0.3 & 1.7 & 1 & \hdots \\
0.1 & 0.2 & 0.1 & 1.3 & 0.5 & \hdots \\
\vdots & \vdots & \vdots & \vdots & \vdots
\end{bmatrix} \\[1em]
\end{split}
\end{equation*}

The objective contribution matrix of the combination length (\(s=3\)) is normalized to create the normalized objective contribution matrix \(\widehat{NC}\) for $s=3$. Summing all $\widehat{NC}$ results for $s\in [1,22]$ values and row-normalizing this $\widehat{S}$ matrix creates the objective contribution result ($\widehat{SC}$) for each site ($i$) and for each objective ($j$).   

\begin{equation*} 
\begin{split}
\widehat{NC}[s=3] & = \begin{bmatrix}
0.08 & 0.14 & 056 & 0.19 & 0.03 & \hdots \\
0.70 & 0.15 & 0.08 & 0.01 & 0.04 & \hdots \\
0.05 & 0.27 & 0.07 & 0.39 & 0.23 & \hdots \\
0.05 & 0.10 & 0.05 & 0.55 & 0.10 & \hdots \\
\vdots & \vdots & \vdots & \vdots & \vdots
\end{bmatrix}
\end{split}
\end{equation*}

\begin{figure}
    \centering
    \includegraphics[width=0.8\textwidth]{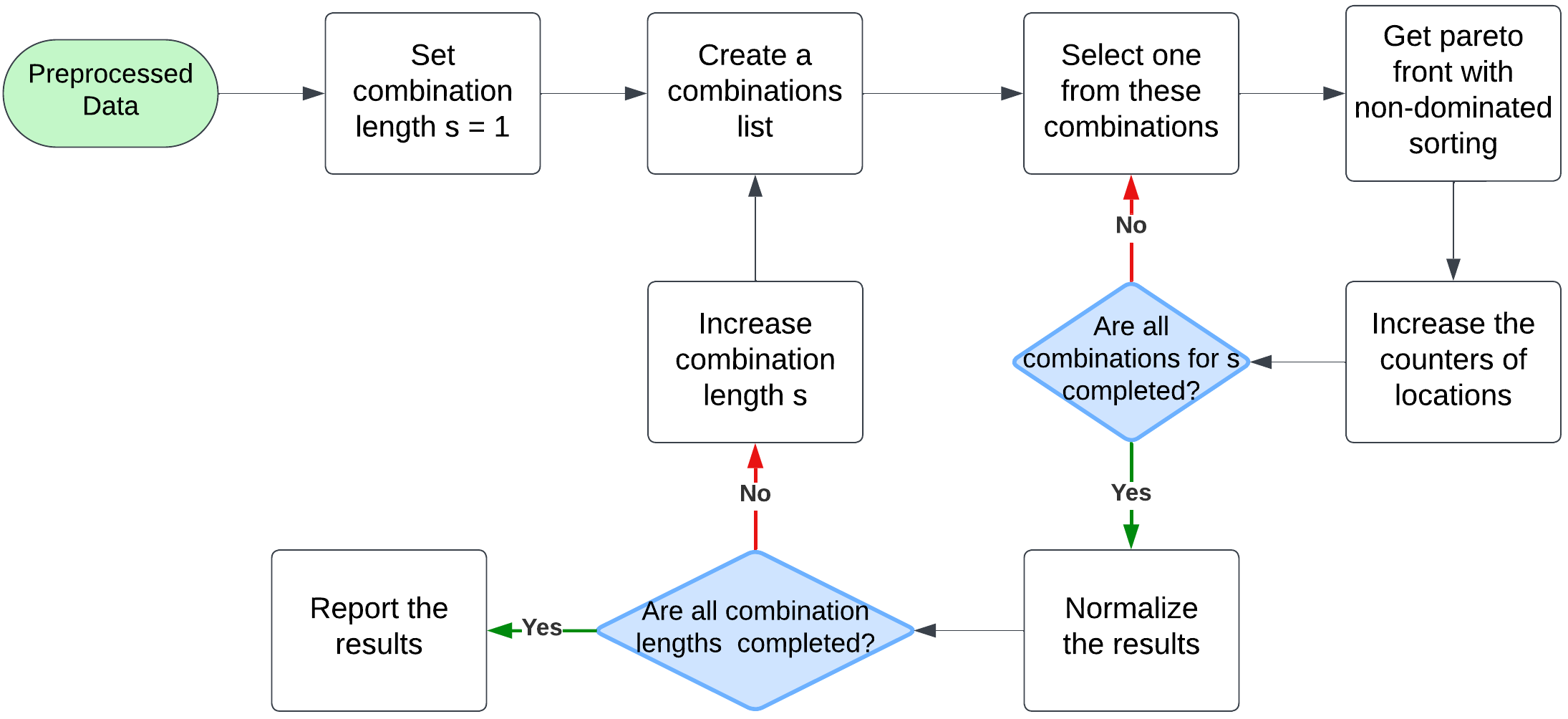}
    \caption{A flowchart of the multi-objective combinatorial analysis for nuclear reactor sites proposed in this study.}
    \label{fig:bf_comb_flowchart}
\end{figure}

We summarize this method in the flowchart shown in Figure \ref{fig:bf_comb_flowchart}. The application of the prescribed method presented several challenges, primarily due to the rapid increase in the number of combinations with varying combination lengths. Prior to the analysis, we executed the siting analysis routine with varying computer core counts and combination lengths. Based on the scaling for core numbers, we anticipated that the complete analysis of all combination lengths would require approximately 5 days running on 360 processors. To mitigate the risk of processes being forcibly terminated due to extended runtimes and to address program memory issues, we implemented a checkpoint system between each combination length. We then executed the method in segments. Each checkpoint records the elapsed computation time for each combination length and the data processed up to that point. We provide a comparison of combination lengths and their corresponding computation times in Table \ref{tab:computation_times}.

\begin{table}[h!]
\caption{Computation times for different combination lengths}
\label{tab:computation_times}
\centering
\begin{center}
\begin{tabular}{|c|c c c c c c c c c|}
\hline
\textbf{Combination Length} & 1 & 2 & 3 & 4 & 5 & 6 & 7 & 8 & 9 \\
\hline
\textbf{Computation Time (min)} & 0.03 & 0.17 & 0.48 & 2.21 & 6.61 & 21.58 & 69.54 & 201.45 & 458.46 \\
\hline
\end{tabular}
\end{center}
\end{table}


\subsection{Concatenated Neural Networks (ConcNN)}
\label{sec:concnn}

To evaluate the impact of each parameter on various locations, we propose a \ac{fnn} model. This model aims to predict the siting metric and the importance of site objectives for any location within the contiguous United States. Additionally, it seeks to identify which objectives will influence siting decisions, eliminating the need to re-run the high-cost analysis described in the previous section. Machine learning applications in nuclear through FNN and other forms of neural networks were broad and include surrogate modeling for expensive reactor simulations \cite{radaideh2020surrogate}, digital twins \cite{song2022online}, multi-objective optimization for nuclear reactor control \cite{price2022multiobjective}, fault detection in nuclear power plants \cite{naimi2022machine}, time series forecasting of nuclear accidents \cite{radaideh2020neural}, among others.  

\hl{The ConcNN model is designed to predict 22 siting objectives, one siting metric, and 22 siting importance values for each site corresponding to each objective. The purpose of creating the neural network model is to decrease the computational time needed for estimating these quantities.} This \ac{fnn} model is composed of two parts. The first part of the model uses the longitude, latitude, county FIPS code (Federal Information Processing Standards\footnote{\url{https://transition.fcc.gov/oet/info/maps/census/fips/fips.txt}}), and state FIPS code as its input and the 22 site objectives as the output. These predicted objectives are then passed to the second stage of the model. In this stage, the 22 objective predictions, along with the original coordinates, county FIPS, and state FIPS, are input to predict the siting metric (score) and the objective importance values. \hl{This training approach reflects the hierarchical relationship where the location influences siting objectives, and siting objectives impact the siting metric and the objective importance values. To replicate this dependency within a unified framework, a single model structure, as depicted in Figure \ref{fig:model_structure}, is implemented for ConcNN, ensuring end-to-end training and prediction within the same model.}

To develop the described model, we defined our  input data ($X$) with four features, including coordinates, state, and county information. This input passes through a series of hidden layers, producing an intermediate output ($Y_1$) that predicts two sets of variables: 14 continuous variables ($Y_L$) and 8 binary variables ($Y_B$). These variables represent the siting objectives, derived from the STAND tool that are relevant to the chosen location. The intermediate output ($Y_1$) is then concatenated with the initial input data ($X$), forming an enhanced feature set. The model then processes this concatenated data through subsequent hidden layers to predict additional siting metrics and objective importance scores ($Y_2$). We then train the model to optimize predictions for both $Y_1$ and $Y_2$ simultaneously. This could improve model performance compared to a standard \ac{fnn} since the initial part of the ConcNN model is further refined using outputs from the second part.

\begin{figure}
    \centering
    \includegraphics[width=0.8\textwidth]{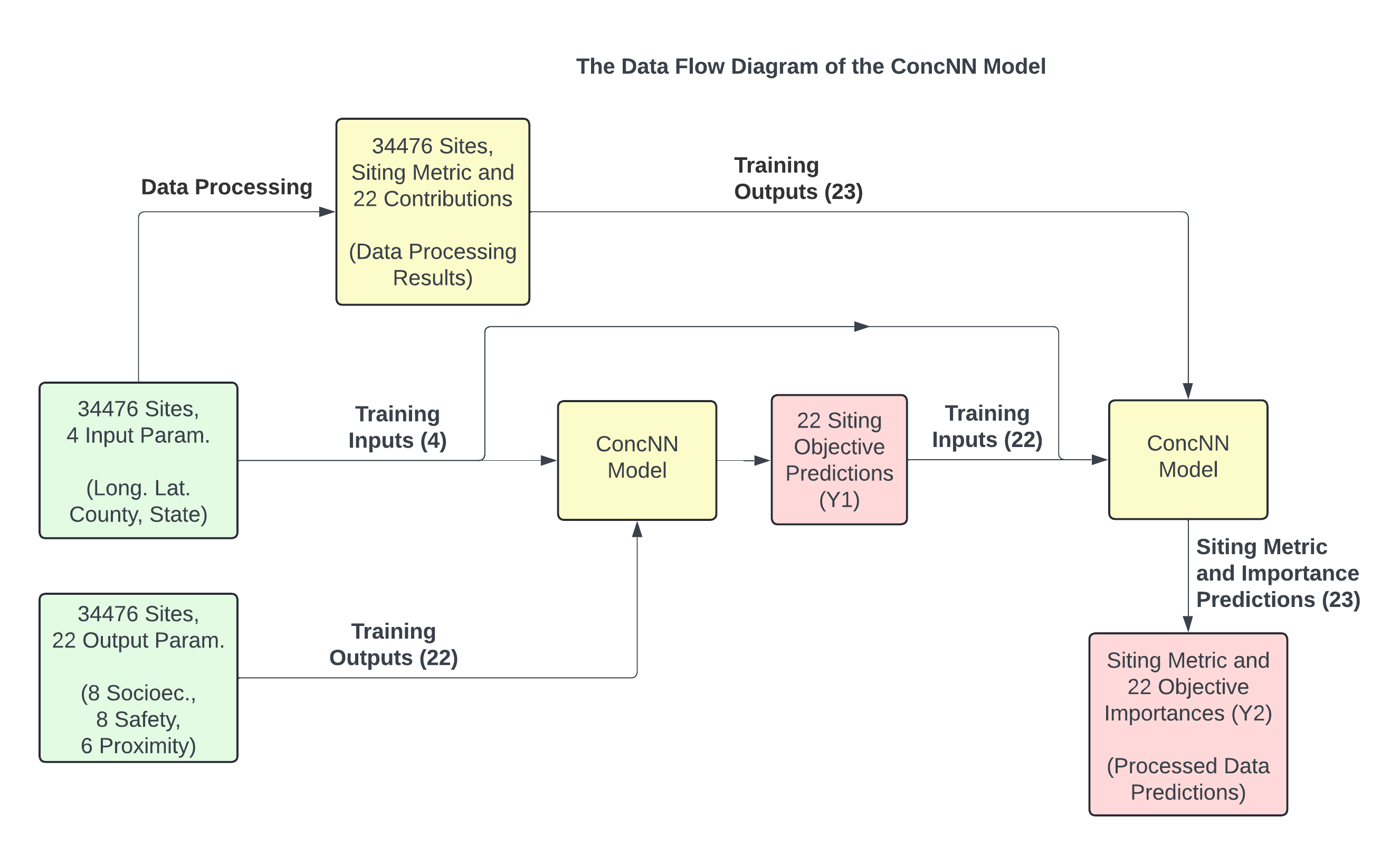}
    \caption{Visualization of the data flow of the ConcNN model (Concatenated Neural Network).}
    \label{fig:ConcNN_Flow}
\end{figure}

We conducted hyperparameter tuning using grid search for each of the following parameters: the number of hidden layers of the first part (ranging from 1 to 8), the number of hidden layers of the second part (ranging from 1 to 8), the number of neurons per layer (between 25 and 1,000), and the learning rate (between 1e-5 and 1e-3). The flowchart illustrating the data flow of the ConcNN model is presented in Figure \ref{fig:ConcNN_Flow}. 

\subsection{Lookup Table Plus Neural Networks (LUT-NN)}
\label{sec:lutnn}

\begin{figure}
    \centering
    \includegraphics[width=0.8\textwidth]{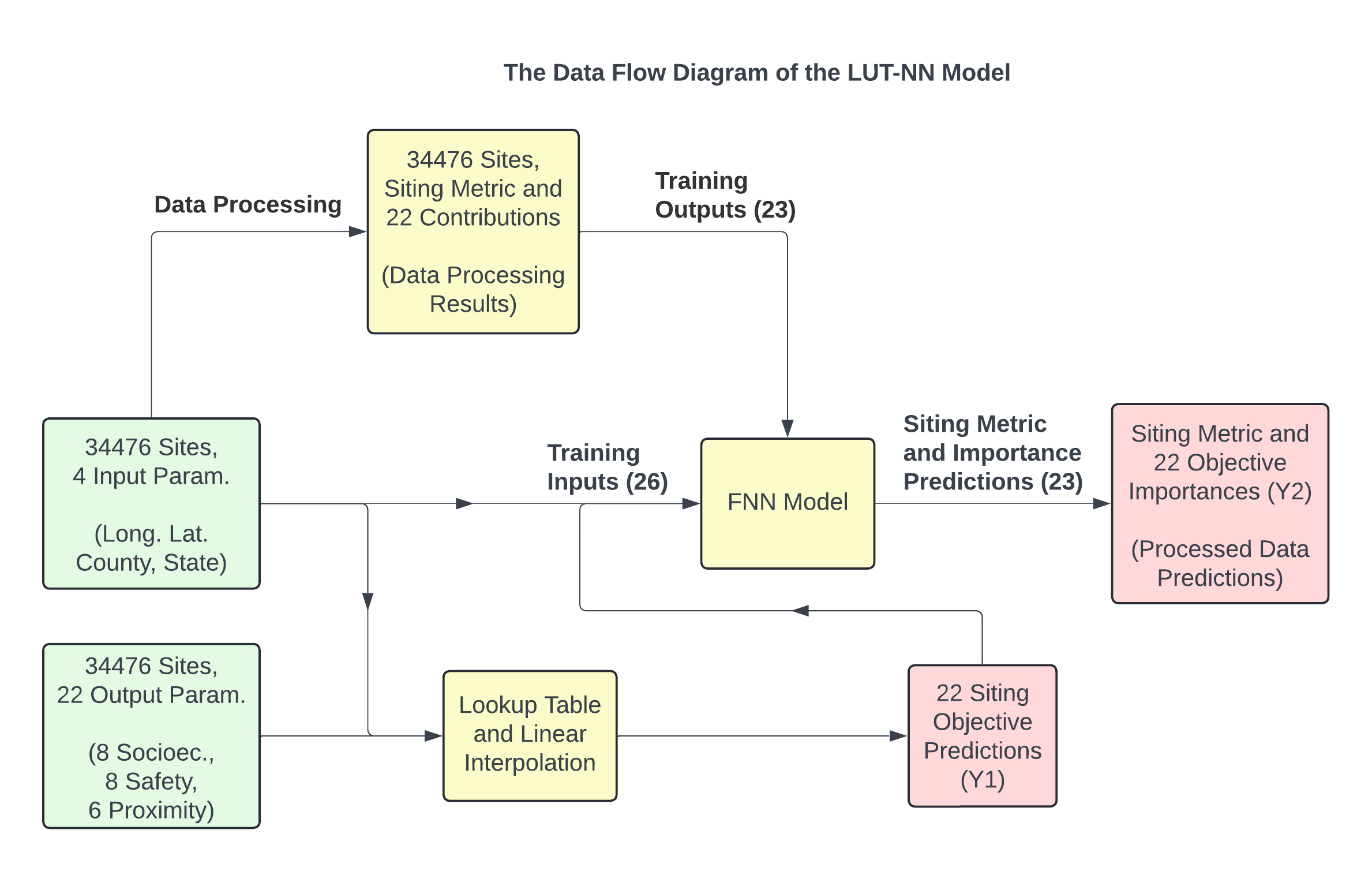}
    \caption{Visualization of the data flow of the LUT-NN model (Lookup Table Plus Neural Networks).}
    \label{fig:LUT-NN_Flow}
\end{figure}

To speed up the calculations and possibly enhance the performance even further, we explored an alternative method to replace the first stage of the ConcNN model. This alternative uses a lookup table combined with a linear interpolation scheme to predict \ac{npp} siting objectives across locations in the contiguous United States. In the original ConcNN model, the first stage processes four input parameters to predict 22 siting objectives. For predicting the characteristics of the locations, the lookup table and interpolation method are expected to perform effectively. This is because most siting objectives are primarily influenced by proximity-based metrics, \hl{with some depending solely on the state and county in which the site is located.} Accordingly, \textbf{LUT-NN} refers to this revised model, which uses a lookup table for predicting the site objectives. \textbf{ConcNN} will be used to refer to the original, fully neural network model that predicts all outputs. Later in the results section, we benchmark both LUT-NN and ConcNN in predicting site objectives, the siting metric, and objective importance values. The data flow schemes for both LUT-NN and ConcNN models are given in Figure \ref{fig:ConcNN_Flow} and \ref{fig:LUT-NN_Flow}, respectively.

\hl{A connection can be established between the proposed neural network models and Explainable Artificial Intelligence (XAI) methods. Various approaches to XAI have been explored, while no single method is universally applicable to all machine learning models. However, the performances of XAI techniques have been well-documented for specific methodologies \cite{leuthe2024leveraging, MACHLEV2022100169}. The research methodology applied in this paper implements a non-dominated multi-objective combinatorial search, designed to aggregate results from each Pareto front where the number of objectives is too large to provide reliable outcomes through traditional non-dominated sorting. This combinatorial search tracks the contribution of each objective to the respective Pareto fronts. By accessing every step of the computation, the methodology ensures transparency, enabling the reconstruction of reasoning for each evaluated combination and demonstrating the influence of individual objectives on the siting metric/score. 

In our computation, we sum contributions from different objectives to a single siting metric, mirroring, for example, the popular SHAP (SHapley Additive exPlanations) approach, where the contribution of each feature in the machine learning model is calculated and aggregated to explain the final prediction \cite{lundberg2017shap}. Additionally, the step where we normalize the objective contributions aligns with SHAP's methodology for distributing credit (or contribution) fairly among objectives \cite{radaideh2019shapley}. SHAP ensures that contributions from all features sum to the predicted outcome (local additivity), similar to our approach. In the context of explainability, our proposed method represents a simplified additive attribution method inspired by SHAP principles.}

\subsection{Computing Resources}
\label{sec:com_res}

\hl{This study used an internal server at the University of Michigan to evaluate the final results of the joint dataset with Brownfield sites and CPP sites as described in Section \ref{sec:comb}. The combinatorial search was performed using a multithreading calculation of 360 threads (180 cores), and successfully completed over a span of five days. For the machine learning training, we employed TensorFlow 2.18.0 and Keras 3.6.0 to develop and train the ConcNN and LUT-NN models. We conducted the training and hyperparameter tuning on the same internal server, which is equipped with two AMD EPYC 9654 processors, each providing 96 cores operating at 2.4–3.7 GHz, resulting in a total of 192 cores and 384 threads. Additionally, the server features four NVIDIA RTX 6000 Ada Generation GPUs and 1536 GB of DDR5 RAM. The training and tuning processes were completed in approximately 18 hours.}

Additionally, the Idaho National Laboratory High-Performance Computing (INL-HPC) system has been used to produce \hl{preliminary results during the early stages of this research before moving to the internal server. This transition occurred due to the restrictions of wall time limits that hindered completing our combinatorial search in an effective manner when a large number of sites were used.} The computations were performed in parallel on the Bitterroot cluster, utilizing 256 cores. Each CPU in the cluster is an Intel® Xeon® Platinum 8480+ with 56 cores operating at a frequency of 3.8 GHz, with 256 GB of RAM per node. 

\section{Results and Discussions}
\label{sec:results}

\subsection{Reactor Siting Analysis in the United States}

\begin{figure}
    \centering
    \includegraphics[width=\textwidth]{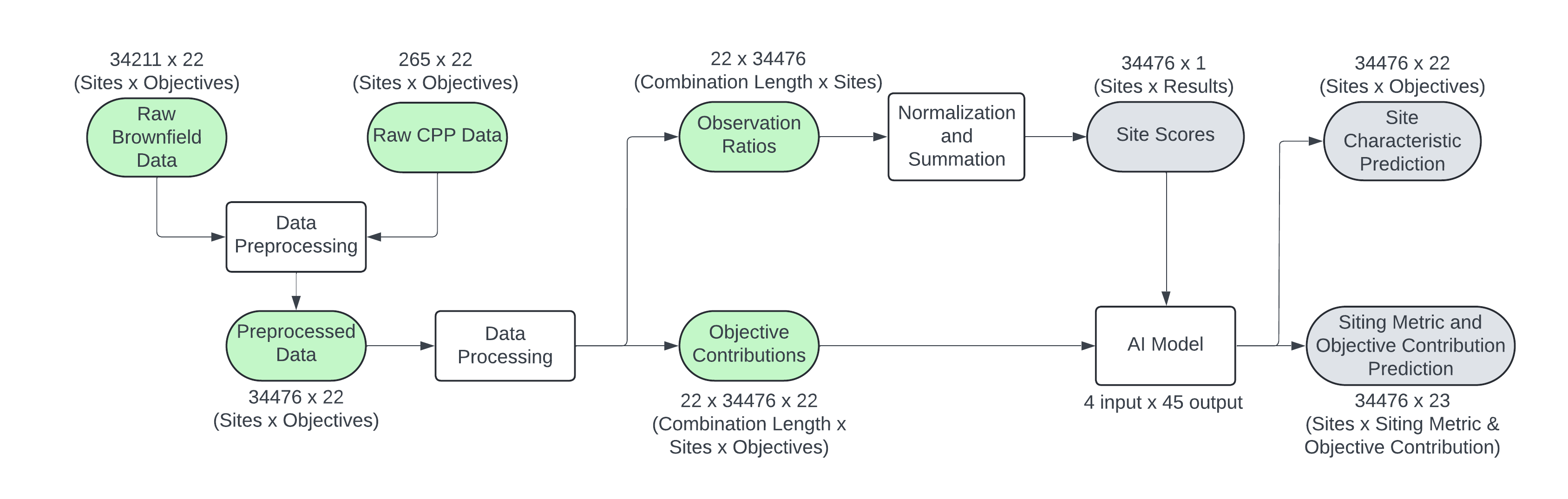}
    \caption{A flowchart summarizing how the site data are combined and processed to feed into the machine learning models}
    \label{fig:data_proc_sum}
\end{figure}

We analyzed the \hl{joint} dataset using the resources explained in Section \ref{sec:com_res} until site observation ratios for all 22 combination lengths were computed based on the method presented in Section \ref{sec:comb}. After this process, we used the resulting data to calculate the siting metric/score. We used this siting score to identify optimal locations and evaluate the influence of each objective on the site score. The workflow employed to derive these findings is summarized in Figure \ref{fig:data_proc_sum}.

\begin{figure}[!h]
    \centering
    \includegraphics[width=0.9\textwidth]{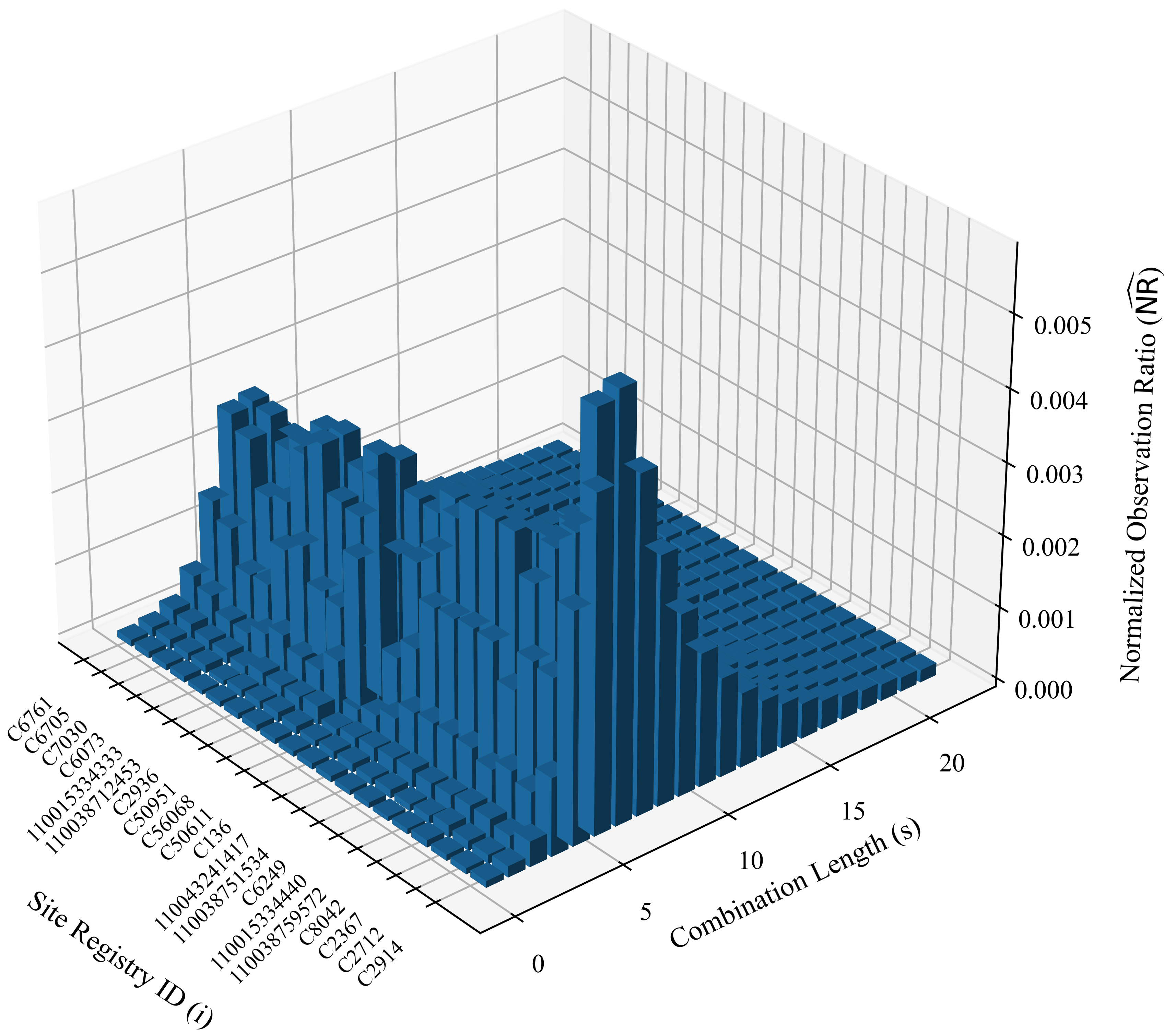}
    \caption{\hl{The normalized observation ratio as a function of combination length from the non-dominated multi-objective combinatorial search. The plot shows the change of $\widehat{NR}$ with combination length ($s$) for the top 20 performing sites.}}
    \label{fig:ground_truth}
\end{figure}

To illustrate how the processed data changes with varying combination lengths, we plot the site observation ratios for different combination lengths prior to summing them. The top-performing locations and their observation ratios across various combination lengths are presented in Figure \ref{fig:ground_truth}. These results indicate that among the top 10 sites, 6 are coal sites and 4 are Brownfield sites. However, in the top 20 ranking, the numbers shift to 14 coal sites and 6 Brownfield sites.

Considering the required computational power, we present the non-dominated, multi-objective combinatorial search operation as a brute-force method for this dataset. We give the product of these operations as a summed and scaled observation ratio (siting score) for each location. The site scores of the best six sites are given in Table \ref{tab:res_explanation}, which include four coal sites and two Brownfield sites. 

\begin{table}[ht]
\caption{\hl{Site scores (summed and scaled observation ratios $\overrightarrow{\mathbb{M}}$) by site ID, state and county for the top six locations in the contiguous United States}}
\label{tab:res_explanation}
\centering
\begin{tabular}{@{}llllll@{}}
\hline
\textbf{Site ID} & \textbf{State} & \textbf{County} & \textbf{Longitude} & \textbf{Latitude} & \textbf{Site Score ($\overrightarrow{\mathbb{M}}$)} \\ \hline
C2914 & Ohio & Tuscarawas & -81.47 & 40.52 & 1.0000 \\ 
C2712 & North Carolina & Person & -79.07 & 36.48 & 0.7991 \\ 
C2367 & New Hampshire & Rockingham & -70.78 & 43.41 & 0.7025 \\ 
C8042 & North Carolina & Stokes & -80.06 & 36.28 & 0.6650 \\ 
110038759572 & Florida & St. Lucie & -80.32 & 27.42 & 0.6169 \\
110015334440 & California & Alameda & -122.10 & 37.67 & 0.6018 \\ \hline
\end{tabular}
\end{table}

\begin{figure}
    \centering
    \includegraphics[width=0.7\textwidth]{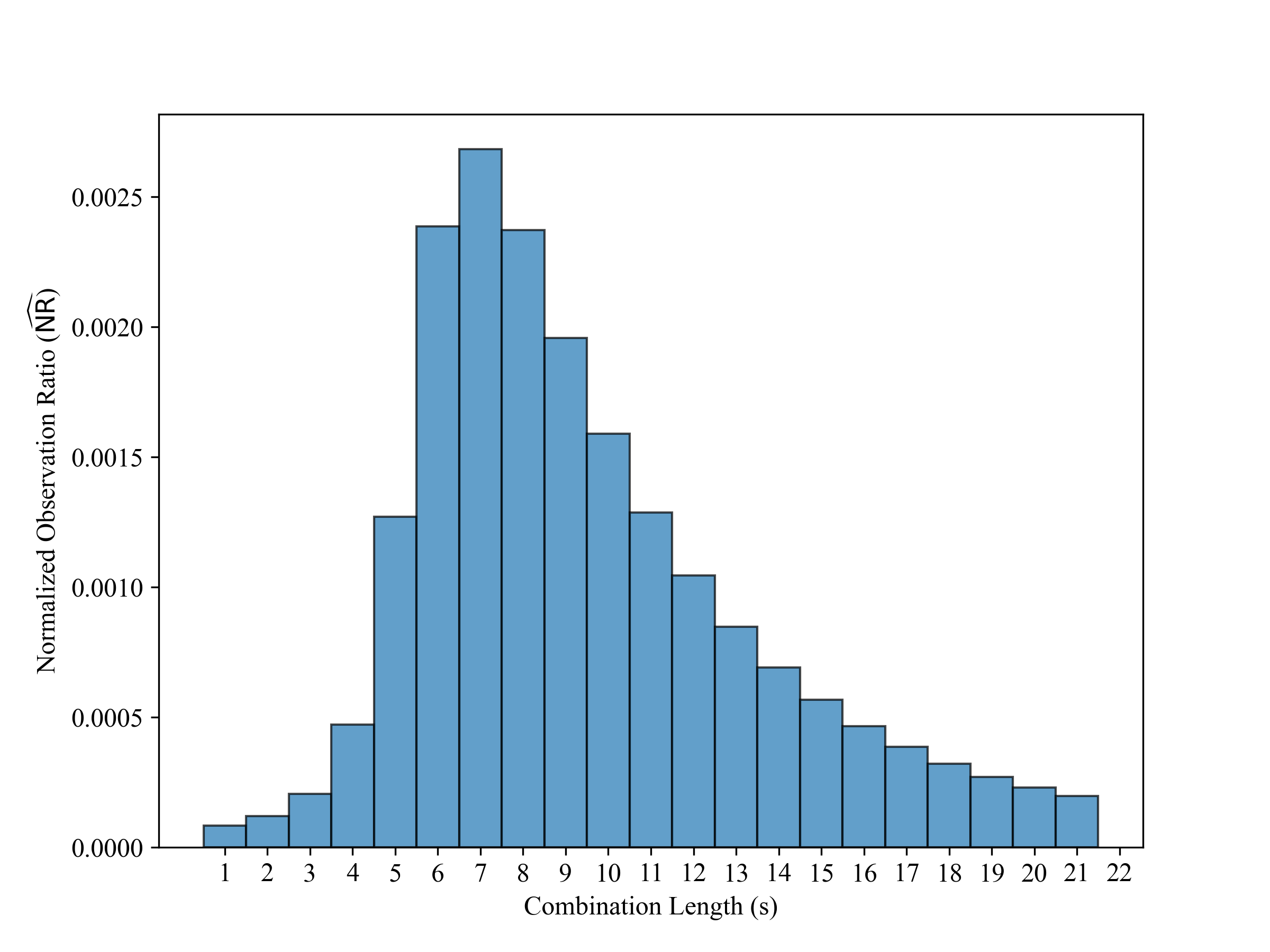}
    \caption{\hl{Normalized observation ratio ($\widehat{NR}$) of the best performing CPP site (CPP ID C2914 in Tuscarawas County, Ohio) for different combination lengths ($s$).}}
    \label{fig:best_ground_truth_2d}
\end{figure}

\begin{figure}
    \centering
    \includegraphics[width=0.7\textwidth]{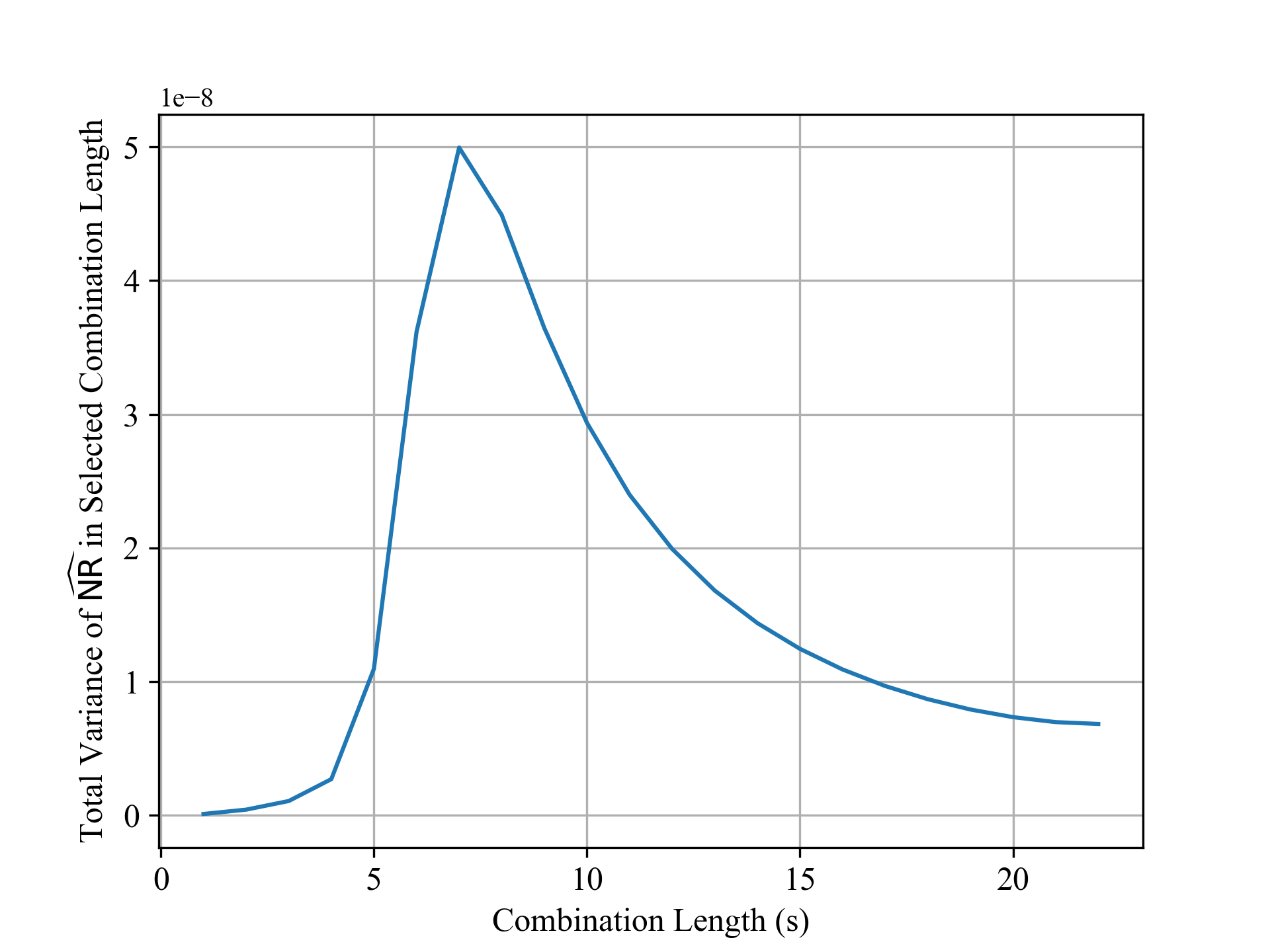}
    \caption{\hl{The variance in the normalized observation ratio as a function of the combination length ($s$). This figure shows how the $\widehat{NR}$ results differ for increasing combination lengths. For $s=1$, the rows of $\widehat{NR}$ are mostly the same (for each site) with approximately 0 variance in the first column of the $\widehat{NR}$ matrix.}}
    \label{fig:variance_analysis}
\end{figure}

For the best-performing site (\hl{CPP ID C2914}, which is located in \hl{Tuscarawas County, Ohio}), the change in the observation ratio is illustrated in Figure \ref{fig:best_ground_truth_2d}. As shown, the observation ratio tends to decrease with increasing combination length. This trend is consistent across all other sites examined in the procedure outlined in Figure \ref{fig:bf_comb_flowchart}. We attribute the observed decrease to the total number of combinations associated with different combination lengths; specifically, fewer combinations at both high and low combination lengths (see Table \ref{tab:combs_explanation}) result in lower observation rates. This becomes more obvious in Figure \ref{fig:variance_analysis}, which shows the change in the normalized observation ratio variance as a function of combination length. Nevertheless, high-order combination lengths (e.g. 18-20) cause larger variance in the observation ratio compared to the low-order ones (e.g., 1-3)  

We show one of the most interesting findings in this study in Figures \ref{fig:ground_truth}-\ref{fig:variance_analysis}, which reveal that the most promising sites tend to exhibit a prominent observation rate on the Pareto front at middle combination lengths, peaking between 6-9 combinations. This suggests that the interaction between objectives plays a critical role in site ranking. Simply assigning weights to individual objectives fails to capture these high-dimensional interactions. These higher-order interactions significantly influence site ranking, while lower-order combination lengths have a relatively smaller impact on the variance of observation rate for different sites.

\hl{As shown in Figure \ref{fig:variance_analysis}, observation ratio variance increases with combination length. For a length of 1, all sites perform nearly identically, but as the length grows, site rankings become more distinct. The highest variance occurs at 7-way combinations, indicating its effectiveness in differentiating sites. For the combination length of 7, the performance disparity is highest between the best and worst sites in the results.  For instance, while lower-order combinations (e.g., 1-3) provide limited separation, higher-order ones (e.g., 8-12) introduce greater variability but require significantly more computation time. While assessing combination lengths with the highest variance would significantly reduce computation time, the most impactful length cannot be determined in advance. As a result, this evaluation cannot be limited to the most effective combination lengths, making it necessary to assess all of them for a thorough site ranking.}

\hl{The top-performing site locations identified in this analysis based on Table \ref{tab:res_explanation} are situated in Tuscarawas County, Ohio (OH); Person County, North Carolina (NC); Rockingham County, New Hampshire (NH); Stokes County, North Carolina (NC), St. Lucie County, Florida (FL); and Alameda County, California (CA). Key distinguishing objectives for these sites include state nuclear inclusive policy, public sentiment toward nuclear energy, number of intersecting protected lands, population center distance, and retiring facility distance. None of these sites has an intersection with a protected land. Of these six sites, only the location in California (6) has a nuclear restriction, which is the need for high-level waste disposal technology or reprocessing capacity. In terms of energy policy and incentives, the sites in Ohio (1) and North Carolina (2, 4) and California (6) have nuclear-inclusive policies, while the site in Florida (5) has highest positive population sentiment towards nuclear energy. The site in Ohio (1) has the lowest substation distance, and has high state electricity imports. The sites in North Carolina (2, 4) has the lowest state construction costs, has all of their safety objectives as positive except the landslide risk, and has no hazardous facilities within 5-miles unlike other sites. Additionally, the top 4 sites have a distance higher than 20 miles to the closest population center. From a safety perspective, the first five sites have most of their safety objectives positive, and are among the safest locations in the whole dataset. The California site (6) is the only site with multiple nearby nuclear R\&D centers. However, it is also very close to hazardous facilities and population centers and does not have a streamflow.}

\begin{figure}[!h]
    \centering
    \includegraphics[width=\textwidth]{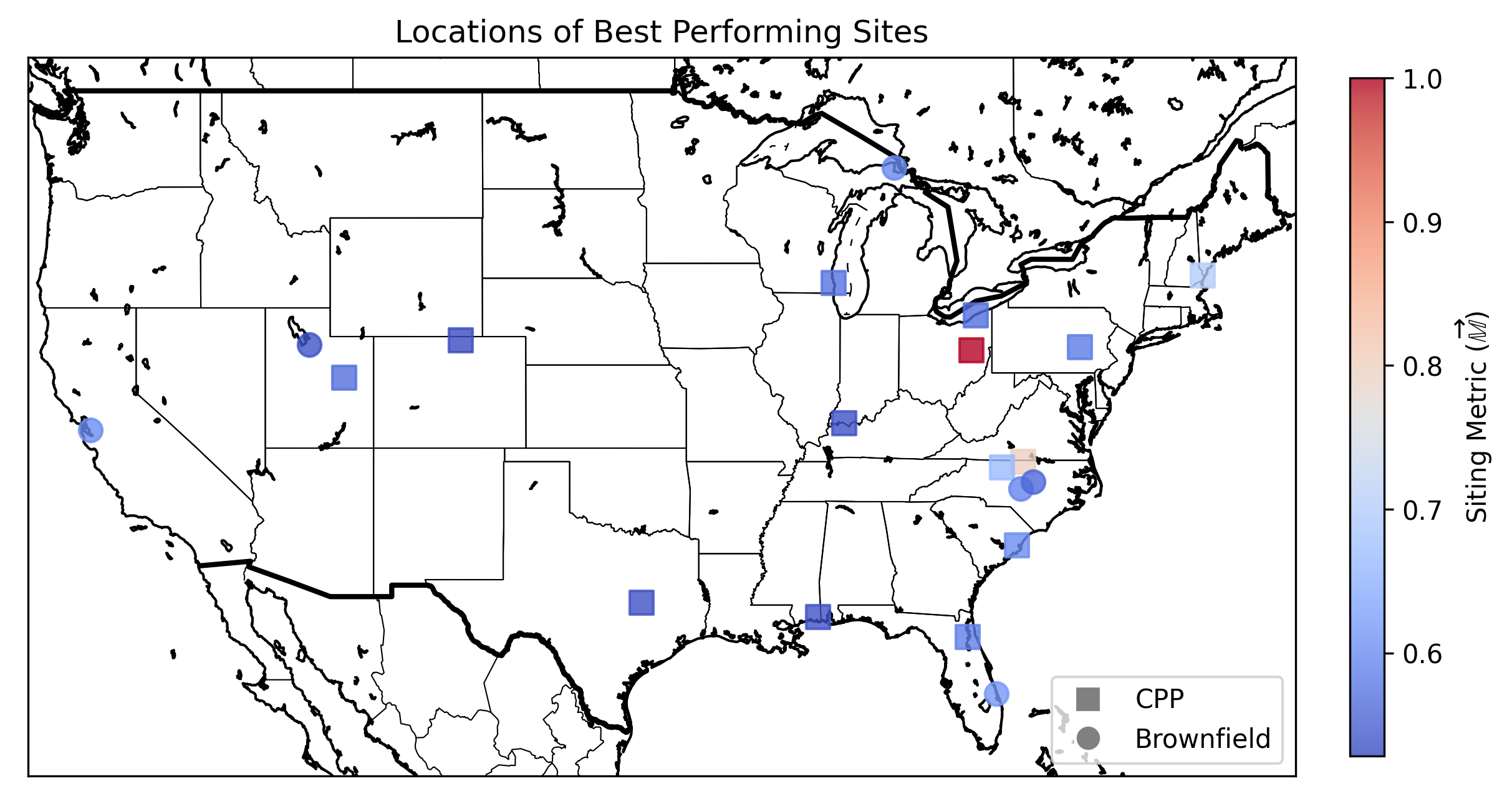}
    \caption{\hl{Best performing coal and Brownfield sites of the joint dataset based on the non-dominated multi-objective combinatorial search.}}
    \label{fig:us_best_comb_map}
\end{figure}

This analysis shows optimal site selection without weighting the importance of specific objectives, thereby yielding an objective conclusion. This is a stark contrast to the subjectivity of status quo methods. The attributes of the top Brownfield and CPP sites highlight the balanced nature of competing objectives in the selection process. The best 20 sites from this comparison, along with their coordinates and site metrics, are illustrated in the map in Figure \ref{fig:us_best_comb_map}. 

\hl{The map identifies six Brownfield sites and 14 CPP sites, each selected based on optimal combinations of 22 objectives from the joint dataset. Upon examining the results of the Brownfield and coal sites, we identify the top four performing sites as CPP locations. While CPPs represent a smaller subset of the overall dataset, their strong siting characteristics contribute to their superior performance. In the top 20 sites, the number of CPP locations are higher than Brownfield locations, meaning that CPP sites are more competitive. Notably, the CPP sites predominantly meet safety objectives, as they are generally located away from external hazards. Coal sites are also strategically positioned near electricity substations, streamflow resources, and major roads, providing an advantage in grid connectivity and infrastructure accessibility over Brownfield sites. Therefore, the prevalence of CPPs among the top 20 selected sites is both expected and justified. While they are not strategically positioned, the Brownfield sites span across all contiguous U.S. states. Given the higher number of Brownfield sites relative to coal power plants, it is expected that some locations comparable to the CPP sites can be identified among Brownfields.}

\begin{figure}
    \centering
    \includegraphics[width=0.9\textwidth]{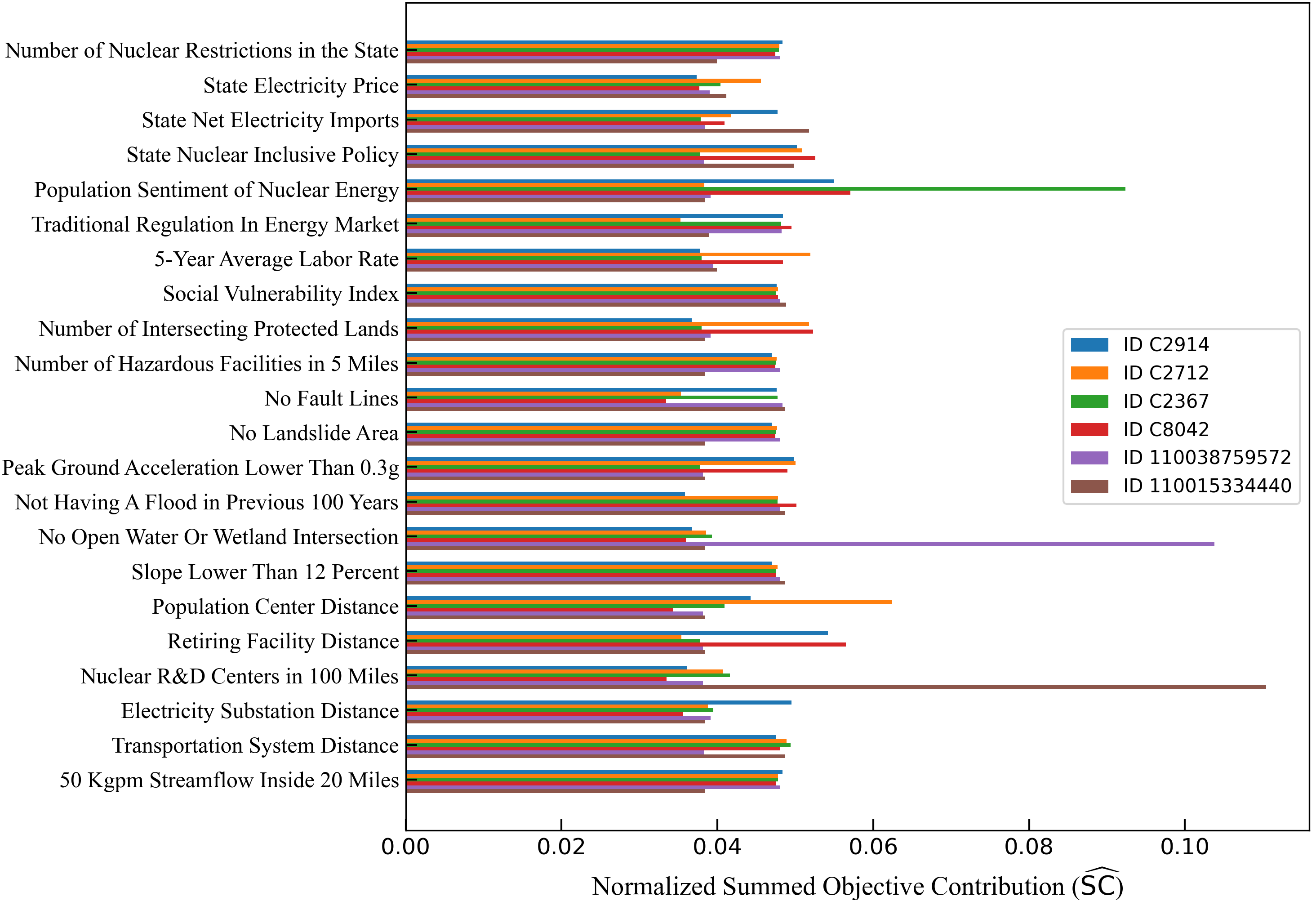}
    \caption{\hl{Summed normalized objective contributions ($\widehat{SC}$) of the best-performing locations in the Brownfield and coal site joint dataset. Further details about each site are given in Table \ref{tab:best_cpp_bf_objectives}.}}
    \label{fig:obj_imp_of_best_bf}
\end{figure}

\hl{The method implemented in this study selects locations that are not dominated by others based on the given criteria. Analysis of objective contributions to the siting score reveals that each site excels in different dominant objectives impacting its performance. The detailed siting score and objective values for the top CPP and Brownfield sites are provided in Table \ref{tab:best_cpp_bf_objectives} in the Appendix. The objective importance values of the best Brownfield and coal sites in the joint dataset are given in Figure \ref{fig:obj_imp_of_best_bf}.}

Our research methodology employs non-dominated sorting to identify locations that dominate across multiple objectives, without considering the specific magnitude of each objective. This approach prioritizes a location’s relative performance on individual objectives rather than their absolute values. Consequently, even if an objective holds the extreme (best or worst) value in the dataset, its specific magnitude is not factored into the sorting process. \hl{For instance, the highest-ranked location has a low electricity price, yet it outperforms others due to advantageous factors such as state electricity imports, safety objectives, proximity to retiring facilities, and distances to electricity substations and population centers.} This outcome underscores that certain important objectives in top locations are outweighed by the sheer number of other objectives. This illustrates how trade-offs become inherent to multi-objective optimization.

\hl{The proposed methodology may unintentionally overlook crucial domain expertise that could better prioritize objectives based on practical constraints, such as safety or cost. For example, if all socioeconomic factors are positive and dominate the dataset, but a single safety objective is extremely low, the methodology might assign a high siting score to this location. However, an expert would quickly recognize that this site is unsuitable for nuclear power plant (NPP) siting. A similar issue arises with the current use of the ``weighted sum method,'' where even if the critical safety objective is assigned a higher-than-average weight, it may still be overshadowed by other objectives with lower-than-average weights, but whose collective impact is still significant. Thus, it is important to note that domain expertise is necessary to validate the results before final site selection, in both the proposed combinatorial method and the weighted sum method. Incorporating domain-specific constraints, hierarchical grouping of objectives, or expert post-hoc validation could help mitigate the risk of an outlier site being ranked too highly.}

\subsection{ConcNN Model Results}

Based on the method presented in Section \ref{sec:concnn} and after collecting an enormous amount of information from our combinatorial search, we built a data-driven model to predict the \ac{npp} siting objectives, siting score, and objective importance values based on the site location. 

\hl{Model evaluation in this section employs various regression metrics including Mean Squared Error (MSE), Root Mean Squared Error (RMSE), Mean Absolute Error (MAE), and $R^2$ (coefficient of determination). MSE measures the average squared difference between predicted and actual values, penalizing larger errors more heavily. RMSE, the square root of MSE, expresses the error in the same units as the target variable, making it more interpretable. MAE calculates the average absolute differences, providing a more direct measure of typical error without emphasizing outliers. $R^2$ evaluates the proportion of variance in the target variable explained by the model, with values closer to 1 indicating better fit. Mathematically, these four metrics are defined as follows:}

\begin{align}
\text{MSE} &= \frac{1}{n} \sum_{i=1}^{n} (y_i - \hat{y}_i)^2, & \quad
\text{RMSE} &= \sqrt{\frac{1}{n} \sum_{i=1}^{n} (y_i - \hat{y}_i)^2}, \\
\text{MAE} &= \frac{1}{n} \sum_{i=1}^{n} |y_i - \hat{y}_i|, & \quad
R^2 &= 1 - \frac{\sum_{i=1}^{n} (y_i - \hat{y}_i)^2}{\sum_{i=1}^{n} (y_i - \bar{y})^2},
\label{eq:metrics}
\end{align}
where $n$ is the number of samples in the test set (e.g., sites not used in training the model), $\hat{y}_i$ is the model prediction of sample $i$ (e.g., site score, objective values, objective importance values), $y_i$ is the true values for sample $i$ that correspond to $\hat{y}_i$ predictions, and $\bar{y}$ is the mean of such true values.

\begin{figure}[!h]
    \centering
    \includegraphics[width=0.85\textwidth]{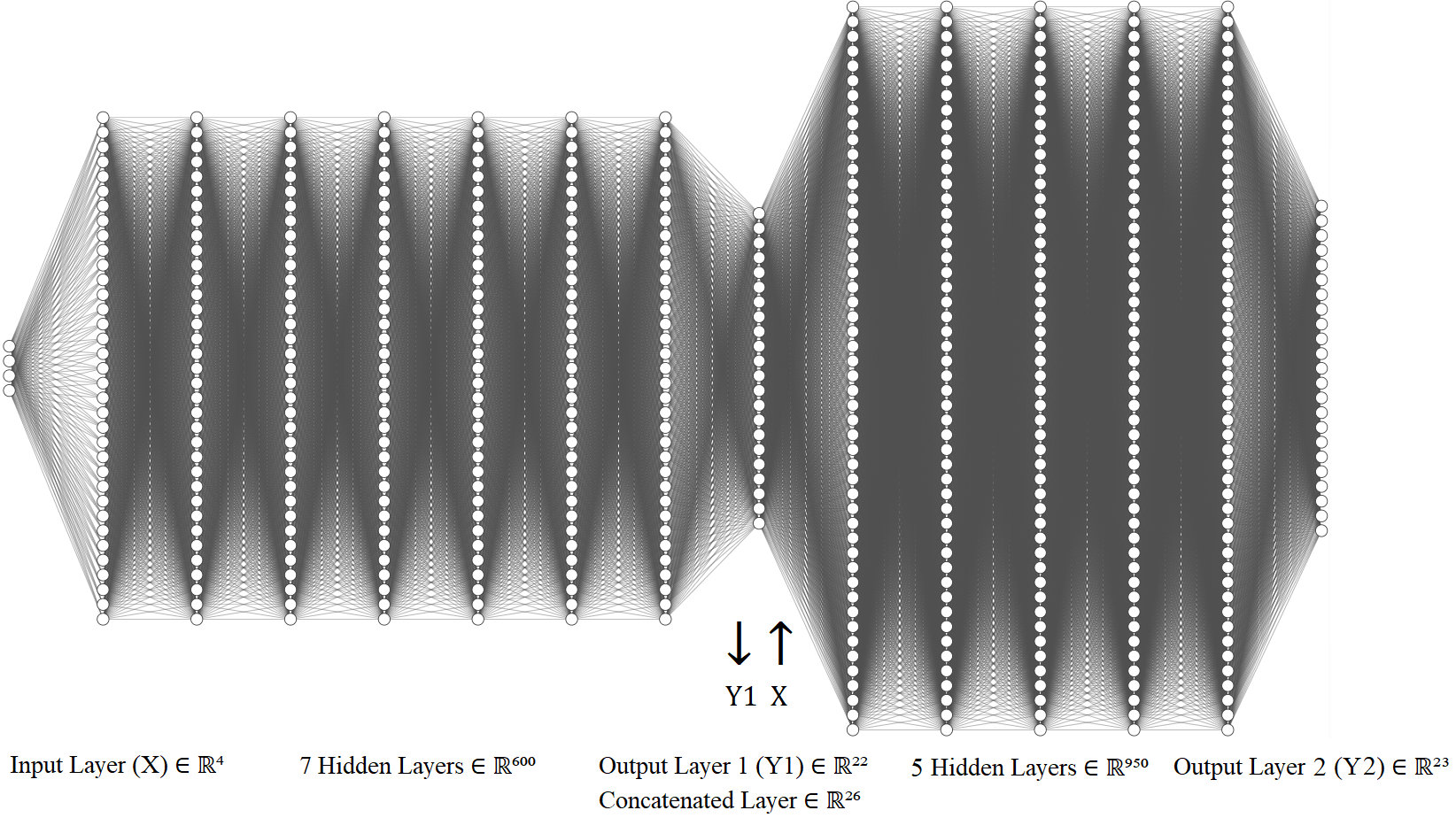}
    \caption{Architecture of the ConcNN model}
    \label{fig:model_structure}
\end{figure}

\begin{figure}[!h]
    \centering
    \includegraphics[width=0.7\textwidth]{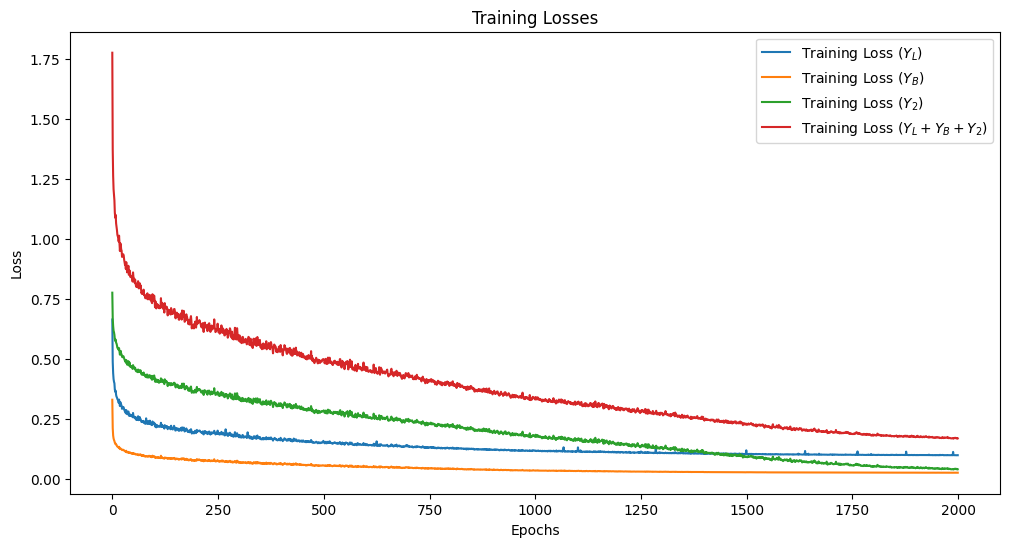}
    \caption{\hl{The change of $Y_L$, $Y_B$, $Y_2$ training losses and the total training loss with the number of epochs for the ConcNN model.}}
    \label{fig:training_losses}
\end{figure}

The best ConcNN model after tuning consists of \hl{seven} layers of \hl{600} neurons in its first part, and \hl{five} layers of \hl{950} neurons in its second part. We set the learning rate to 1e-3. We select the test data proportion to be 20\% of the total dataset (\hl{6,895 site locations}). The test set is used to assess model performance through site locations not used during training. We set the number of epochs to 2,000 and the batch size to 256. The siting objective prediction ($Y1$) and siting score and objective importance prediction ($Y2$) training metrics for the ConcNN model are given in Table \ref{tab:model_metrics}. The model training loss is given in Figure \ref{fig:training_losses}. The best ConcNN architecture is sketched in Figure \ref{fig:model_structure}.

\begin{table}[ht] 
\centering
\caption{\hl{ConcNN Model Training and Test Metrics}}
\label{tab:model_metrics}
\begin{tabular}{ccccc}
\hline
\textbf{Metric} & \textbf{MSE} & \textbf{RMSE} & \textbf{MAE} & \textbf{R²} \\
\hline
 \text{Y1 Training} & 66485 & 257.848 & 44.8439 & 0.86388  \\
\text{Y1 Test} & 101831 & 319.110 & 51.3142 & 0.77205  \\
\hline
\text{Y2 Training} & 0.00028 & 0.00911 & 0.00313 & 0.92680  \\ 
\text{Y2 Test} & 0.00027 & 0.01654 & 0.00662 & 0.75634  \\
\hline
\end{tabular}
\end{table}

The metric results in Table \ref{tab:model_metrics} indicate that while the ConcNN model performs reasonably well, it lacks high accuracy in predicting site characteristics, with an $R^2$ on the test set ranging from 0.75 to 0.75 for its two output layers ($Y1$ and $Y2$), compared to the ideal $R^2$ of 1.0. This suggests that, despite being trained on a dataset of approximately 34,000 points, the problem remains challenging to model purely by machine learning. The primary difficulty arises because the model receives only 4 input features related to the site's location and is tasked with predicting 22 site objectives. As a result, predictions may be particularly error-prone near state borders where site objectives may exhibit significant variability in response to minor changes in input parameters. Furthermore, the lower accuracy in predicting $Y1$ propagates to $Y2$, leading to even lower $R^2$ values for the second output. Despite these limitations, the authors believe that the model can still offer rapid predictions of site objectives, site scores, and objective importance values that capture general trends across U.S. sites. However, its errors may be significant enough to limit its utility for precise site ranking. 

\subsection{LUT-NN Model Results}

In this section, we present an assessment of LUT-NN, which replaces the first part of ConcNN with a lookup table and linear interpolation. The hyperparameter tuning for the LUT-NN was conducted over a broad range of values: the number of layers varied from 1 to 10, the number of neurons per layer ranged from 25 to 1,000, and the learning rate was tested between 1e-3 and 1e-5. \hl{During the hyperparameter tuning, a parameter grid has been generated for the defined parameter range. The best model was selected after the grid search was completed over all possible combinations of the parameters. After hyperparameter tuning, the optimal model configuration identified consisted of five layers with 950 neurons in each layer, and an initial learning rate of 2e-4. We applied a learning rate decay factor of 0.92 with early stopping patience set to 25 epochs and trained the model for 1,000 epochs with a batch size of 16.} To address overfitting, we applied L2 regularization with values ranging from 1e-2 to 1e-5, as well as dropout rates between 0.05 and 0.5, to each layer. Unfortunately, these strategies were not effective in improving prediction accuracy for locations that are absent from the lookup table. 

We present the prediction metrics for LUT-NN on the $Y1$ values (22 site objectives) in Table \ref{tab:interpolation_metrics} for test sites that already ``exist'' in the dataset. As expected, the model performs perfectly for these sites, reflected by the metrics in Table \ref{tab:interpolation_metrics}, as the model only needs to match the input data with the right site index. \hl{However, the interpolation results heavily depend on the locations used in the lookup table. In order to assess the $Y1$ metrics for site locations not present in the lookup table, a 100-fold test set has been generated with 1\% of the existing lookup table as test data and the remaining as the interpolation data.} Using this separated test data, as shown in Table \ref{tab:interpolation_metrics}, the averaged performance of LUT-NN in interpolating siting objectives was found to be comparable to the predictions made by the first part of the ConcNN model ($Y1$). Specifically, the test $R^2$ is approximately 0.77 for the first part of ConcNN and 0.80 for LUT-NN. This result suggests that the first stage of the ConcNN model does not capture more information than what can be achieved through linear interpolation between sites.

\begin{table}[ht] 
\centering
\caption{\hl{Interpolation test metrics for LUT-NN for the first set of outputs (e.g., Y1, 22 site objectives)}}
\label{tab:interpolation_metrics}
\begin{tabular}{ccccc}
\hline
\textbf{Metric} & \textbf{MSE} & \textbf{RMSE} & \textbf{MAE} & \textbf{R²} \\
\hline
\text{Existing Site} & 0.00000 & 0.00000 & 0.00000 & 1.00000  \\
\text{Objective ($Y_1$) Interpolation (Non-existing Site)} & 17.52051 & 4.06095 & 0.69053 & 0.80116  \\
\hline
\end{tabular}
\end{table}

\hl{The test metrics (MSE, RMSE, and MAE) for interpolation in Table \ref{tab:interpolation_metrics} differ significantly from those of the first-layer ConcNN (Y1). The interpolation table retrieves existing state information when site data is requested. Since this lookup table contains data from all contiguous U.S. states, it perfectly predicts state-dependent attributes. These attributes include key socioeconomic characteristics: (1) the number of nuclear restrictions in the state, (2) state electricity price, (3) state net electricity imports, (4) state nuclear-inclusive policy, (6) traditional regulation in the energy market, and (7) the 5-year average labor rate. Among these, (3) state net electricity imports and (7) the 5-year average labor rate have values that are considerably larger in magnitude compared to the other characteristics. Because the first-layer ConcNN exhibits errors in these columns while the lookup table does not, the MSE, RMSE, and MAE metrics between these methods differ significantly.}

The most notable difference between the two models emerges when the site objectives ($Y1$) are used to predict the second set of outputs ($Y2$: site score and objective importance values). As noted earlier in Table \ref{tab:model_metrics}, the $Y2$ metrics for ConcNN are even lower than those for $Y1$ with 20\% test sites withheld from the full dataset. However, as shown in Table \ref{tab:model2_metrics}, when LUT-NN is used to predict $Y2$ for the same test sites as ConcNN, we see significant improvement. The training $R^2$ increases to approximately 0.97, while the test $R^2$ improves to 0.84 for \hl{the locations close to the lookup table sites} and 0.79 for \hl{the locations far to the lookup table sites}, compared to 0.75 for ConcNN. \hl{This demonstrates that using LUT-NN to predict site scores and objective importance values is considerably more accurate than relying solely on a neural network-based approach. As previously mentioned, we applied various regularization techniques to mitigate overfitting, and the results in Table \ref{tab:model2_metrics} represent the best outcomes achieved.}

\begin{table}[ht] 
\centering
\caption{\hl{LUT-NN Training and Test Metrics when predicting $Y_2$ based on interpolated $Y_1$ values}}
\label{tab:model2_metrics}
\begin{tabular}{ccccc}
\hline
\textbf{Metric} & \textbf{MSE} & \textbf{RMSE} & \textbf{MAE} & \textbf{R²} \\
\hline
\text{$Y_2$ Training} & 0.000027 & 0.005201 & 0.000878 & 0.973305 \\
\text{$Y_2$ Test (Without Y1 Interpolation Error)} & 0.000162 & 0.012728 & 0.004583 & 0.840029 \\
\text{$Y_2$ Test (With Y1 Interpolation Error)} & 0.000237 & 0.015380 & 0.004566 & 0.79412 \\
\hline
\end{tabular}
\end{table}

For both ConcNN and LUT-NN, the user is required only to input the location coordinates along with the FIPS codes for the county and state, which are available from the Federal Communications Processing System\footnote{\url{https://transition.fcc.gov/oet/info/maps/census/fips/fips.txt}}. The model then predicts NPP siting objectives, siting metrics, and objective importances by interpolating between more than 34,000 site locations available to the model. We designed these models to facilitate accessible and efficient retrieval of \ac{npp} siting information across locations in the contiguous United States. Based on our analysis, ConcNN can be used for general site assessment without ranking, but the LUT-NN is much more reliable when predicting both site objectives for any site in the United States. LUT-NN should be preferred for site ranking, as the site scores and objective importance values are more accurate than those predicted by ConcNN.  

\hl{One key advantage of these data-driven models, along with their rapid prediction capabilities, is their ability to support sensitivity analysis and uncertainty quantification—both of which require repeated model executions. These tasks are impractical with combinatorial search methods, as the process must be repeated whenever an objective is removed. However, with the LUT-NN model, users can exclude an objective (e.g., electricity substation distance) and assess its impact on the site score using techniques like one-at-a-time sensitivity analysis. Alternatively, they can conduct more comprehensive variance-based sensitivity analyses, such as those using Sobol indices, while keeping computational costs manageable. To maintain the focus of this work, we defer in-depth sensitivity analysis and uncertainty quantification to future studies.}

\hl{The development of this machine learning-based NPP site selection algorithm has profound implications for the future of nuclear energy and infrastructure planning. By using data-driven decision-making built on data generated by unbiased combinatorial search, we minimize human bias, accelerate site selection, and enhance the overall process. The development of this method may help in receiving regulatory approvals, optimizing the costs of site selection, and ensuring that nuclear sites are strategically located for grid stability and resource efficiency. Additionally, it fosters public trust by promoting transparency in the selection process. On a global scale, this approach can aid both developed and emerging nations in efficiently identifying optimal sites for NPPs. Ultimately, this research paves the way for a more comprehensive evaluation of nuclear energy infrastructure crucial for meeting future energy demands and climate goals.}


\section{Concluding Remarks}
\label{sec:conc}

This study introduces a novel, multi-objective combinatorial methodology for \ac{npp} site assessment and ranking that eliminates the need for analyst-defined weights, reducing potential bias. The study marks the first comprehensive evaluation of a vast number of potential nuclear reactor sites in the United States, considering both Brownfield and coal sites, using a unified, flexible methodology. The proposed approach is adaptable and can be applied to other countries beyond the United States. Furthermore, we developed a machine learning model using the extensive dataset generated through this combinatorial process that enables rapid assessments of nuclear reactor site suitability across the United States. This model requires only basic site location information (site coordinates, county, and state) from the user. 

We generated a comprehensive dataset encompassing raw siting objectives for \ac{npp}s in the United States for both Brownfield and coal sites. Researchers can use this dataset in its entirety (22 objectives) or can focus on specific objectives relevant to their studies. It includes a wide range of socioeconomic, safety, and proximity factors. Notably, most of these siting characteristics are uncorrelated, preserving distinct information about various locations across the United States. 

\hl{In this analysis, we identified some specific Brownfield sites that present viable opportunities for the siting of NPPs even when compared with the CPP sites.} The findings, detailed previously, demonstrate that these specific sites possess the necessary socioeconomic, geographic, environmental, and proximity characteristics, and have advantages in terms of existing infrastructure and minimized land-use conflicts. Future research should continue to explore the implications of \ac{npp} siting on Brownfield sites, specifically considering other factors, such as community acceptance to ensure holistic decision-making. To ensure the thorough evaluation and justification of the final siting decisions for the proposed \ac{npp}s, \hl{we expect that a detailed socio-techno-economic analysis is needed for the top-ranked sites found in this study including both the coal and Brownfield sites.}

\hl{Our analysis also reveals a clear situation in the comparison of nuclear energy generation performance between CPPs and Brownfield sites for NPP siting. While some Brownfields demonstrate comparable performance compared to CPPs, most CPP sites outperform the entire Brownfield dataset. This finding underscores the justification of the selected CPP sites when considered in the context of energy site assessment. This conclusion agrees with the previous research conducted in this area \cite{Rafi,Hansen_inv} using methods distinct from those in this study. Given these findings, it is also shown that some exceptional Brownfields can compete with CPPs when socioeconomic, safety, and proximity characteristics are adequately considered. The findings of this research indicate that Brownfields should be considered alongside coal sites for nuclear projects.}

\hl{One limitation observed in this work is that neural network models still have room for improvement, particularly in their ability to generalize to new sites. This challenge likely stems from data variability and model overfitting. The current dataset may be insufficient to ensure broad generalization across all states, highlighting the need for more data to enhance accuracy and robustness. Future research should explore advanced regularization techniques and improved data acquisition methods to address these issues. Additionally, further work on data-driven models should include sensitivity analysis and uncertainty quantification to assess their robustness against both minor and major perturbations in input data.}


\section*{Data Availability}
\label{sec:avail}

The dataset and models created in this study are available on the public Github repository: \url{https://github.com/aims-umich/NPP_Siting}.

\section*{Acknowledgment}
This work is sponsored by the Department of Energy Office of Nuclear Energy's Distinguished Early Career Program (Award number: DE-NE0009424), which is administered by the \ac{neup}. This research also made use of Idaho National Laboratory computing resources for generating preliminary results, which are supported by the Office of Nuclear Energy of the U.S. Department of Energy under Contract No. DE-AC07-05ID14517. 

\section*{CRediT Author Statement}

\begin{itemize}
    \item \textbf{Omer Erdem}: Methodology, Software, Validation, Formal analysis, Visualization, Investigation, Writing - Original Draft. 
    \item \textbf{Kevin Daley}: Conceptualization, Data Curation, Writing - Review and Edit.  
    \item \textbf{Gabriel Hoelzle}: Conceptualization, Data Curation, Writing - Review and Edit.  
    \item \textbf{Majdi I. Radaideh}: Conceptualization, Methodology, Funding acquisition, Supervision, Resources, Project administration, Writing - Review and Edit.
\end{itemize}

\clearpage
\appendix
\begin{minipage}{\textwidth}
\section{Site Characteristics of the Top Coal and Brownfield Sites}
\begin{table}[H]
\centering
\scriptsize 
\caption{Result table of the best Brownfield and \ac{cpp}s found in this analysis.}
\begin{tabular}{|p{2.5cm}!{\vrule width 1.5pt}c|c|c|c|c|c|}
\hline
\textbf{Registry ID} & C2914 & C2712 & C2367 & C8042 & 110039000000 & 110015000000 \\
\hline
\textbf{Type} & Coal & Coal & Coal & Coal & Brownfield & Brownfield \\
\hline
\textbf{Longitude} & -81.4682 & -79.0731 & -70.7842 & -80.0603 & -80.32 & -122.10 \\
\hline
\textbf{Latitude} & 40.5201 & 36.4833 & 43.0978 & 36.2811 & 27.42 & 37.67 \\
\hline
\textbf{County \& State} & Tuscarawas, OH & Person, NC & Rockingham, NH & Stokes, NC & St. Lucie, FL & Alameda, CA \\
\hline
\textbf{Siting Metric} & 1.0000 & 0.7991 & 0.7025 & 0.6650 & 0.6169 & 0.6018 \\
\hline
\textbf{State Nuclear Restrictions} & 0 & 0 & 0 & 0 & 0 & 1 \\
\hline
\textbf{State Electricity Price} & 11.9544 & 14.0500 & 12.6467 & 12.1533 & 13.3900 & 24.2231 \\
\hline
\textbf{State Net Electricity Imports} & 37952 & 14875 & -6623 & 14875 & 13226 & 75504 \\
\hline
\textbf{State Nuclear Inclusive Policy} & 1 & 1 & 0 & 1 & 0 & 1 \\
\hline
\textbf{Population Sentiment of Nuclear Energy} & 0.0741 & 0.4586 & 0.0300 & 0.0904 & 0.5901 & 0.4019 \\
\hline
\textbf{Traditional Regulation In Energy Market} & 1 & 0 & 1 & 1 & 1 & 0 \\
\hline
\textbf{5-Year Average Labor Rate} & 43920 & 30384 & 38740 & 30384 & 31532 & 48022 \\
\hline
\textbf{Social Vulnerability Index} & 0.4111 & 0.4266 & 0.4347 & 0.4139 & 0.5058 & 0.3835 \\
\hline
\textbf{Number of Intersecting Protected Lands} & 0 & 0 & 0 & 0 & 0 & 0 \\
\hline
\textbf{Number of Hazardous Facilities in 5 Miles} & 2 & 0 & 5 & 0 & 2 & 88 \\
\hline
\textbf{No Fault Lines} & 1 & 1 & 1 & 1 & 1 & 0 \\
\hline
\textbf{No Landslide Area} & 1 & 0 & 1 & 0 & 1 & 1 \\
\hline
\textbf{Peak Ground Acceleration Lower Than 0.3g} & 1 & 1 & 1 & 1 & 1 & 0 \\
\hline
\textbf{Not Having A Flood in Previous 100 Years} & 1 & 1 & 0 & 1 & 0 & 0 \\
\hline
\textbf{No Open Water Or Wetland Intersection} & 0 & 1 & 1 & 1 & 1 & 1 \\
\hline
\textbf{Slope Lower Than 12 Percent} & 1 & 1 & 1 & 1 & 1 & 1 \\
\hline
\textbf{Population Center Distance} & 132.511 & 209.498 & 68.441 & 23.753 & 2.458 & 1.577 \\
\hline
\textbf{Retiring Facility Distance} & 0 & 205.292 & 78.225 & 0 & 32.546 & 3.142 \\
\hline
\textbf{Nuclear R\&D Centers in 100 Miles} & 0 & 1 & 2 & 0 & 0 & 66 \\
\hline
\textbf{Electricity Substation Distance} & 0.0269 & 0.2077 & 0.1330 & 0.2162 & 0.1530 & 1.6407 \\
\hline
\end{tabular}
\label{tab:best_cpp_bf_objectives}
\end{table}
\end{minipage}

\clearpage
\bibliographystyle{elsarticle-num}
\setlength{\bibsep}{0pt plus 0.3ex}
{
\bibliography{references}}

\end{document}